\documentclass[lettersize,journal]{IEEEtran}
\usepackage[utf8]{inputenc} 
\usepackage[T1]{fontenc}
\usepackage{url}
\usepackage{ifthen}
\usepackage[cmex10]{amsmath}
\usepackage{array}

\usepackage{enumerate}
\usepackage{amssymb}
\usepackage{amsmath} 

\usepackage{booktabs}
\usepackage{multirow}

\usepackage{algorithm}
\usepackage{algorithmicx}
\usepackage{algpseudocode} 
\usepackage{subfig}
\usepackage{cite}

\usepackage{threeparttable}

\usepackage{mathrsfs}
\usepackage{bm}
\usepackage{tikz}
\usetikzlibrary{arrows}
\usepackage{graphicx,booktabs,multirow}
\usepackage{epstopdf}
\usepackage{multirow}
\usepackage{tabularx} 
\usepackage{booktabs}
\usepackage[font={small}]{caption}

\usepackage{pifont}

\definecolor{colorhkust}{RGB}{20,43,140}
\definecolor{colortsinghua}{RGB}{116,52,129}
\definecolor{color1}{RGB}{128,0,0}

\usepackage{amsthm}
\usepackage{cite}
\newtheorem{proposition}{Proposition}
\newtheorem{remark}{Remark}

\theoremstyle{definition}
\newtheorem{definition}{Definition}

\theoremstyle{remark}

\newcommand{\diagg}{\mathrm{diag}}

\begin{document}

        \title{Task-Oriented Communication for Multi-Device Cooperative Edge Inference}

      \author{Jiawei~Shao,~\IEEEmembership{Graduate Student Member,~IEEE,}
        Yuyi~Mao,~\IEEEmembership{Member,~IEEE,}
        and~Jun~Zhang,~\IEEEmembership{Fellow,~IEEE}
      	\thanks{J. Shao and J. Zhang are with the Department of Electronic and Computer Engineering, The Hong Kong University of Science and Technology, Hong Kong (E-mail: jiawei.shao@connect.ust.hk, eejzhang@ust.hk). Y. Mao is with the Department of Electronic and Information Engineering, Hong Kong Polytechnic University, Hong Kong (E-mail: yuyi-eie.mao@polyu.edu.hk). This work was supported by the General Research Fund (Project No. 15207220) and the Research Impact Fund (R5009-21) from the Hong Kong Research Grants Council. The work of Y. Mao was supported by a start-up fund of the Hong Kong Polytechnic University under Grant P0038174. (The corresponding author is J. Zhang.)}

}
\maketitle
\begin{abstract}
This paper investigates task-oriented communication for multi-device cooperative edge inference, where a group of distributed low-end edge devices transmit the extracted features of local samples to a powerful edge server for inference.
While cooperative edge inference can overcome the limited sensing capability of a single device, it substantially increases the communication overhead and may incur excessive latency. 
To enable low-latency cooperative inference, we propose a learning-based communication scheme that optimizes local feature extraction and distributed feature encoding in a \textit{task-oriented} manner, i.e., to remove data redundancy and transmit information that is essential for the downstream inference task rather than reconstructing the data samples at the edge server.
Specifically, we leverage Tishby's information bottleneck (IB) principle \cite{tishby2000informationIB} to extract the task-relevant feature at each edge device, and adopt the distributed information bottleneck (DIB) framework of Aguerri-Zaidi \cite{DVIB} to formalize a single-letter characterization of the optimal rate-relevance tradeoff for distributed feature encoding.
To admit flexible control of the communication overhead, we extend the DIB framework to a distributed deterministic information bottleneck (DDIB) objective that explicitly incorporates the representational costs of the encoded features. 
As the IB-based objectives are computationally prohibitive for high-dimensional data, we adopt variational approximations to make the optimization problems tractable.
To compensate for the potential performance loss due to the variational approximations, we also develop a selective retransmission (SR) mechanism to identify the redundancy in the encoded features among multiple edge devices to attain additional communication overhead reduction.
Extensive experiments on multi-view image classification and multi-view object recognition tasks evidence that the proposed task-oriented communication scheme achieves a better rate-relevance tradeoff than existing methods.
\end{abstract}


% \fi
\IEEEpeerreviewmaketitle

%\IEEEPARstart{T}{his} demo file is intended to serve as a ``starter file''
%for IEEE Communications Society journal papers produced under \LaTeX\ using
%IEEEtran.cls version 1.8b and later.
%I wish you the best of success.

\section{Introduction}

The recent revival of artificial intelligence (AI) has affected a plethora of application domains, including computer vision \cite{computer_vision}, augmented/virtual reality (AR/VR) \cite{vr}, and natural language processing (NLP) \cite{natureLP}.
Most recently, as edge devices (e.g., mobile phones, sensors) are becoming increasingly widespread and sensory data are easy to access, the potential of AI technologies has been exemplified at the network edge, which is a new research area named \emph{edge inference} \cite{shi2020communication}, \cite{shao2020communication}, \cite{reviewer_Moldoveanu1}.
Typical edge inference systems convey all the input data from the edge devices to a physically close edge server (e.g., a base station) and leverage deep neural networks (DNNs) to perform inference.
However, as the volume of the collected data (e.g., images, high-definition videos, and point clouds) is enormous in typical mobile intelligent applications \cite{self-driving_cars}, \cite{unlu2019deep_drone}, \cite{zou2019collaborative_robotics}, directly transmitting the raw data to the edge server may lead to a prohibitive communication overhead.

To enable low-latency inference, \emph{device-edge co-inference} stands out as a promising solution \cite{zhou2019edge_edge_intell}, \cite{iccshao}, \cite{shao2020branchy}, which extracts task-relevant features from the raw input data and forwards them to be processed by an edge server. 
This exemplifies a recent shift in communication system design for emerging mobile applications, named \emph{task-oriented communication}, which optimizes communication strategies for the downstream task \cite{shao2021learning}. 
While device-edge co-inference, together with task-oriented communication, has enabled low-latency inference for resource-constrained devices, its performance is limited by the sensing capability of a single device. 
In many applications, e.g., smart drones \cite{unlu2019deep_drone}, robotics \cite{zou2019collaborative_robotics}, and security camera systems \cite{surveillance}, cooperative inference among multiple devices can significantly enhance the inference performance. 
Nevertheless, the communication overhead and latency may also increase substantially with the number of devices, forming the bottleneck for multi-device cooperative edge inference.
Although most existing works on task-oriented communication have led to efficient communication for single-device edge inference, simply reusing these methods for multi-device cooperative edge inference is infeasible.
This is because the correlation among the extracted features of different devices cannot be effectively exploited, which leads to excessive redundancy and latency in transmission.
To address the limitation, this paper will develop
task-oriented communication strategies for multi-device cooperative edge inference by leveraging Tishby's information bottleneck (IB) principle \cite{tishby2000informationIB}, as well as the distributed information bottleneck (DIB) framework proposed by Aguerri and Zaidi \cite{DVIB}.

% derived from distributed source coding theories {\color{red} Inaki Estella Aguerri et al.} \cite{courtade2013multiterminal}.

\subsection{Related Works and Motivations}

The recent line of research on task-oriented communication \cite{strinati20216g_goal_oriented}, \cite{zhu2020toward} has motivated a paradigm shift in communication system design. %\emph{data-oriented communication} towards task-oriented communication.
The conventional \emph{data-oriented communication} aims to guarantee the correct reception of every single transmitted bit.
However, with the increasing number of edge devices, the limited communication resources at the wireless network edge cannot cope with the request to transmit the huge volume of data.
The task-oriented communication provides an alternative scheme to handle this challenge, which regards a part of the raw input data (e.g., nuisance) as irrelevant and considers to extract and transmit only the task-relevant feature that could influence the inference result.
As recovering the original input data with high fidelity at the edge server is unnecessary, task-oriented communication schemes leave abundant room for communication overhead reduction.
{This insight is consistent with the information bottleneck (IB) principle \cite{tishby2000informationIB}: A good extracted feature should be \emph{sufficient} for the inference task while retaining \emph{minimal} information from the input data.
Thus, extracting a \emph{minimal sufficient feature} in task-oriented communication improves the transmission efficiency.}

Recently, there exist several studies taking advantages of deep learning to develop effective feature extraction and encoding methods following the task-oriented design principle \cite{shao2020branchy}, \cite{shao2021learning}, \cite{bottlenet}, \cite{dubois2021lossy}.
In particular, for the image classification task, an end-to-end architecture was proposed in \cite{bottlenet} that leverages a JPEG compressor to encode the internal features extracted by the on-device neural network.
In contrast to data-oriented communication that concerns the data recovery quality, the proposed method is directly trained with the cross-entropy loss for the targeted classification task. 
Similar ideas were utilized in the image retrieval task \cite{jankowski2020wireless_Jankowski} and the point cloud processing application \cite{shao2020branchy}.
Besides, the authors of \cite{dubois2021lossy} characterized the minimum bit rate of the extracted features required to ensure satisfactory performance of the prediction tasks.
These features are learned from the unsupervised neural compressors that bound the encoded rates.
{Although the aforementioned works can largely reduce the communication overhead in single-device edge inference, simply extending them to the multi-device scenario is not efficient as their coding schemes cannot fully exploit the correlation among the extracted features of different devices, which necessitates the investigation of distributed feature encoding.

Distributed coding is an important problem in information theory and communication, which regards the data encoding of multiple correlated sources from different devices that do not communicate with each other.
%Remarkably, the Slepian-Wolf theorem \cite{slepian1973noiseless_SW_Theorem} showed that the distributed coding rate is asymptotically {\color{red}equal to the joint coding rate.}
%Remarkably, the Slepian-Wolf theorem \cite{slepian1973noiseless_SW_Theorem} showed that the distributed coding rate is asymptotically {\color{red}equal to the joint coding rate.} for correlated memoryless sources.
Remarkably, the Slepian-Wolf Theorem \cite{slepian1973noiseless_SW_Theorem} shows that distributed coding can achieve the joint coding rate for correlated memoryless sources.}
In other words, the redundancy among the dependent sources can be identified and eliminated before transmission.
This classic result has recently received renewed attention and led to several works \cite{hanna2020distributed}, \cite{singhal2020communication}, \cite{choi2019context}, \cite{Multi-Robot_Collaborative_Percep_GNN} to develop DNN-based distributed feature encoding schemes in task-oriented communication.
The authors of \cite{hanna2020distributed} developed a distributed quantization system tailored to the classification task, and proposed a greedy search algorithm as part of the deep learning approach to create rectangular quantization regions.
Besides, a significance-aware feature selection mechanism was proposed in \cite{singhal2020communication}, which leverages a DNN model to determine the significance value of each feature and communicates only the features that are likely to impact the inference result.
These methods have shown effectiveness in saving communication bandwidth compared with the single-device edge inference.
However, there lacks a theoretical and rigorous way to characterize the tradeoff between the communication cost and the inference performance in distributed feature encoding, hindering further performance enhancement.

Recently, the distributed information bottleneck (DIB) framework \cite{DVIB} was proposed to investigate multi-view learning, which establishes a single-letter characterization of the optimal tradeoff between the rate (or the representational cost) and relevance (or task-relevant information) for a class of memoryless sources using mutual information terms.
%seeks the right balance between data fit and generalization by using the mutual information as both a cost function and a regularizer.
Such an information-theoretical principle fits nicely with the multi-device cooperative edge inference as it can provide an optimal distributed coding scheme under the assumption that the input views are conditionally independent of the inference variable.
In this paper, we consider a general scenario where the input views may not be conditionally independent and add a feature extraction step before the coding process to improve the communication efficiency.

\subsection{Contributions}
% The proposed task-oriented communication schemes for multi-device cooperative edge inference adds an extra feature extraction step to discard the task-irrelevant correlated information across input views.
% We add an extra feature extraction step to discard the task-irrelevant correlated information across input views.
% Besides, piratical distributed feature encoding schemes have been proposed that takes the finite-blocklength effect into consideration.
% The task-oriented communication schemes for multi-device cooperative edge inference
% In this paper, we develop effective task-oriented communication schemes for multi-device cooperative edge inference based on the IB principle and the DIB framework.

In this paper, we develop task-oriented communication strategies for cooperative edge inference, based on the IB principle for task-relevant feature extraction and the DIB framework for distributed feature encoding.
The major contributions are summarized as follows:
\begin{itemize}
\item  We formulate two design problems for task-oriented communication in multi-device cooperative edge inference systems, namely task-relevant feature extraction and distributed feature encoding.
Specifically, we leverage the IB principle to design the local feature extractor, which aims at discarding the task-irrelevant correlated information across input views.
%which aim at discarding the redundant information from the raw input data.
%Our theoretical results prove that the minimal sufficient features satisfy the conditional independence, which allows us to apply the DIB framework to formalize the optimal rate-relevance tradeoff in the distributed feature encoding.
%Besides, the DIB framework is adopted to formalize the optimal rate-relevance tradeoff in the {\color{red} asymptotic limit}.
Besides, to obtain a practical one-shot coding scheme, we reformulate the DIB objective as the distributed deterministic information bottleneck (DDIB) objective that explicitly incorporates the representational costs of the encoded features (i.e., the number of bits to be transmitted).
%Besides, the DIB framework is adopted to formalize a single-letter characterization of the optimal rate-relevance tradeoff in the distributed feature encoding, which is reformulated as the distributed deterministic information bottleneck (DDIB) objective to incorporate the representational costs of the encoded features.
% {\color{gray}
% \item  We formulate two design problems for task-oriented communication in multi-device cooperative edge inference systems, namely task-relevant feature extraction and distributed feature encoding.
% Specifically, we leverage the IB principle to design the task-relevant feature extraction at each edge device, which aims at extracting the \emph{minimum sufficient} features from the raw input data.
% Our theoretical results prove that the minimal sufficient features satisfy the conditional independence, which allows us to apply the DIB framework to formalize the optimal rate-relevance tradeoff in the distributed feature encoding.
% To obtain a practical one-shot coding scheme, we reformulate the DIB objective as the distributed deterministic information bottleneck (DDIB) objective that explicitly incorporates the representational costs of the encoded features (i.e., the number of bits to be transmitted).
% }
\item Observing that the mutual information terms in the IB-based objectives are computationally prohibitive for high-dimensional data, we resort to the variational approximations to devise tractable upper bounds for the IB and DDIB objectives, resulting in the variational information bottleneck (VIB) objective and the variational distributed deterministic information bottleneck (VDDIB) objective, respectively.
\item To compensate for the potential performance loss due to the variational approximations, we introduce a selective retransmission (SR) mechanism in the VDDIB framework to identify the redundancy among the encoded features, which helps to further reduce the communication load. {It also provides a flexible way to trade off the communication overhead with the inference performance.}
\item The effectiveness of the proposed task-oriented communication strategies is validated on multi-view image classification tasks and multi-view object recognition tasks, {which demonstrates a substantial reduction in the communication overhead compared with the traditional data-oriented communication strategy}.
Moreover, extensive simulation results demonstrate that our methods outperform existing learning-based feature encoding methods for cooperative edge inference.
\end{itemize}

\subsection{Organization}

The rest of this paper is organized as follows. 
Section \ref{Sec:system_model} introduces the system model and formulates the problems of task-relevant feature extraction and distributed feature encoding based on the IB and DIB frameworks, respectively.
Section \ref{Sec:Variational_feature_extraction} introduces the VIB objective as a variational upper bound of the IB formulation for tractable optimization.
Section \ref{Sec:Variational_Distributed_Coding} reformulates the DIB objective to the DDIB objective and derives a tractable VDDIB objective for optimization of the distributed feature encoding.
%Section \ref{Sec:Selective_Transmission} introduces the 
A selective retransmission mechanism is developed in Section \ref{Sec:Selective_Transmission} to further reduce the communication overhead.
In Section \ref{Sec:Experiment}, we provide extensive simulation results to evaluate the performance of the proposed methods. 
Finally, Section \ref{Sec:Conclusion} concludes the paper.

\subsection{Notations}

Throughout this paper, upper-case letters (e.g. $X,Y$) and lower-case letters (e.g. $\bm{x,y}$) stand for random variables and their realizations, respectively. 
We use $X_{1:K}$ to represent the set of random variables $(X_{1},\ldots,X_{K})$.
The entropy of $Y$ and the conditional entropy of $Y$ given $X$ are denoted as $H(Y)$ and $H(Y|X)$, respectively. The mutual information between $X$ and $Y$ is represented as $I(X,Y)$, and the Kullback-Leibler (KL) divergence between two probability distributions $p(\bm{x})$ and $q(\bm{x})$ is denoted as $D_{\mathrm{KL}}\left(p||q\right)$.
The statistical expectation of $X$ is denoted as $\mathbf{E}\left(X\right)$.
We further denote the Gaussian distribution with mean vector $\bm{\mu}$ and covariance matrix $\bm{\Sigma}$ as $\mathcal{N}\left(\bm{\mu},\bm{\Sigma}\right)$, and use $\bm{I}$ to represent the identity matrix.
The element-wise product is denoted as $\odot$.

\section{System Model and Problem Description}
\label{Sec:system_model}

\begin{figure}[t]
    \centering
    \includegraphics[width=1\linewidth]{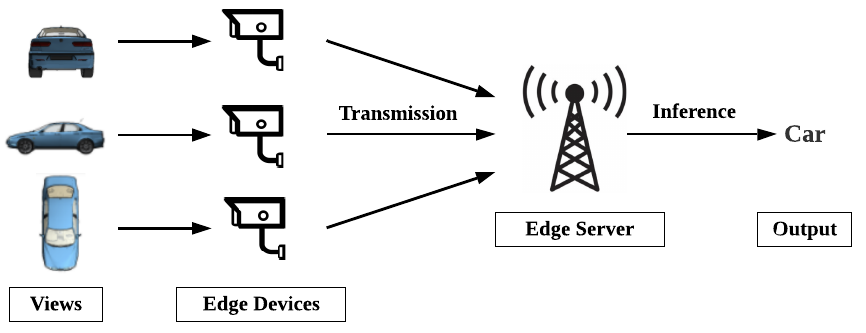}
    \caption{An example of multi-device cooperative edge inference for multi-view object recognition.}
    \label{Fig:distributed_system_example}
\end{figure}

We consider a multi-device cooperative edge inference system, where a group of low-end edge devices collaborate to perform an inference task with the assistance of an edge server.
These devices perceive different views of the same object, and therefore cooperative inference can effectively overcome the limited sensing capability of a single device \cite{su2015multi_MVCNN}.
Related application scenarios include multi-view object detection \cite{su2015multi_MVCNN} in security camera systems and vehicle re-identification \cite{surveillance} in urban surveillance.
A sample system for multi-view object recognition is shown in Fig. \ref{Fig:distributed_system_example}.
Compared with the edge devices, the edge server has abundant computational resources so that server-based inference can be executed in real time by collecting the sensing data from multiple edge devices.
Nevertheless, edge inference systems typically suffer from limited communication bandwidth, and thus directly transmitting the raw sensory data (e.g., high-definition images and point clouds) from the edge devices to the edge server shall incur significant communication latency.
To enable low-latency cooperative edge inference, it is critical to design efficient task-oriented communication strategies that only transmit the task-relevant information to the edge server for device-edge co-inference.
Next, we first present a probabilistic modeling of the multi-device cooperative edge inference systems and introduce two key design problems of task-oriented communication, namely, task-relevant feature extraction and distributed feature encoding.

\subsection{Probabilistic Modeling}

\begin{figure}[t]
    \centering
    \includegraphics[width=1\linewidth]{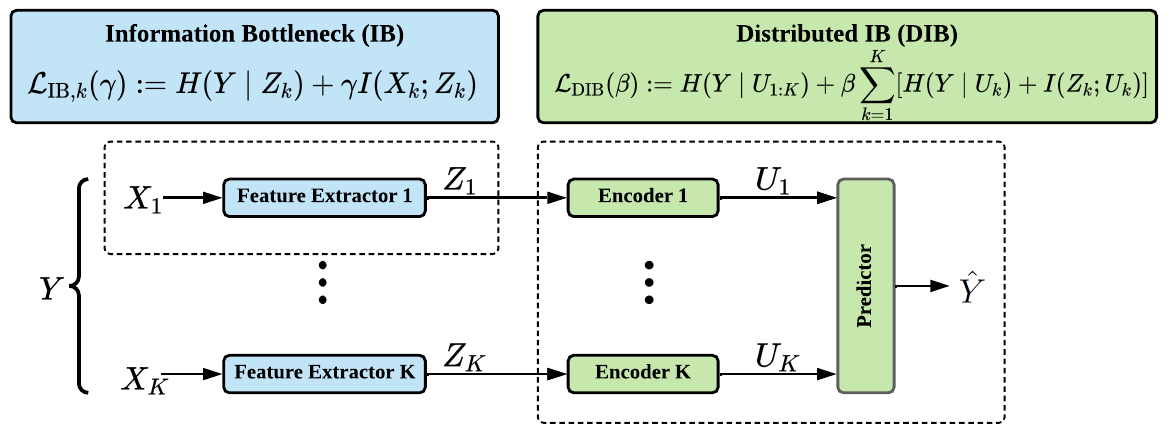}
    \caption{Probabilistic modeling for multi-device cooperative edge inference systems.}
    \label{Fig:distributed_system}
\end{figure}

The probabilistic modeling for multi-device cooperative edge inference, comprising $K$ edge devices and an edge server, is shown in Fig. \ref{Fig:distributed_system}.
The input views of different edge devices, denoted as $ (\bm{x}_{1},\ldots,\bm{x}_{K})$, and the target variable $\bm{y}$ (e.g., the category of an object), are deemed as different realizations of the random variables $(X_{1},\ldots,X_{K},Y)$ with joint distribution $p(\bm{x}_{1},\ldots,\bm{x}_{K},\bm{y})$.
%$P_{X_{1:K},Y}({X_{1},\ldots,X_{K},Y})$. 
To perform cooperative inference, each edge device first extracts the task-relevant feature $\bm{z}_{k}$ from the input $\bm{x}_{k}$, and encodes the extracted feature to a compact vector $\bm{u}_{k}$ for efficient transmission to the edge server.
After receiving the encoded features from the edge devices, the edge server performs further processing to derive the final inference result.
The extracted features $(\bm{z}_{1},\ldots,\bm{z}_{K})$ and the encoded features $(\bm{u}_{1},\ldots,\bm{u}_{K})$ are instantiated by random variables $(Z_{1},\ldots,Z_{K})$ and $(U_{1},\ldots,U_{K})$, respectively.
These random variables constitute the following Markov chain:
\begin{equation}
\label{equ:probability_graphical_model}
    Y \leftrightarrow X_{k} \leftrightarrow Z_{k} \leftrightarrow U_{k}, \ \forall k \in \{1,\ldots,K\},
\end{equation}
which satisfies $p(\bm{u}_{k},\bm{z}_{k},\bm{x}_{k}|\bm{y}) = p(\bm{u}_{k}|\bm{z}_{k})p(\bm{z}_{k}|\bm{x}_{k})p(\bm{x}_{k}|\bm{y})$, $\forall k \in \{1,\ldots,K\}$.
A key design objective is to minimize the communication overhead while providing satisfactory inference performance. The proposed solution involves effective local feature extraction and distributed feature encoding, which are specified in the following two subsections.


\subsection{Task-Relevant Feature Extraction}
\label{Subsec:IB}

Given the high dimension of the sensing data, a significant part of the input $X_k$ at the $k$-th device is superfluous to the target variable $Y$.
To achieve efficient task-oriented communication, the extracted feature $Z_{k}$ should capture only the task-relevant information about the target variable $Y$ from the observation $X_{k}$.
% {\color{blue}
% The feature extraction step would benefit the next distributed feature encoding step.
% The extracted features can be less correlated since the task-irrelevant correlated information could be discarded.
% }
Our design follows the information bottleneck (IB) principle \cite{tishby2000informationIB}, which was shown to be effective for feature extraction and encoding in single-device edge inference systems \cite{shao2021learning}.
The core idea is to maximize the mutual information between the encoded feature $Z_{k}$ and the target variable $Y$ while minimizing the mutual information between the encoded feature $Z_{k}$ and the input view $X_{k}$.
The objective function is formulated as follows:
\begin{equation}
\label{equ:IB_extract_final_obj}
\mathcal{L}_{\mathrm{IB},k}(\gamma) := H(Y|Z_{k}) + \gamma I(X_{k};Z_{k}),
\end{equation}
where $\gamma \geq 0$ is a weighting factor, and the constant term $H(Y)$ is eliminated as $I(Y;Z_{k}) = H\left(Y\right) - H\left(Y|Z_k\right)$.
As minimizing the IB objective function requires high computation cost to obtain the marginal distribution of $Z_{k}$, we adopt the variational method \cite{DVIB} to reformulate (\ref{equ:IB_extract_final_obj}) into an amenable form in Section \ref{Sec:Variational_feature_extraction}.

\begin{remark}
\label{remark:minimal_sufficient_feature}
The feature extraction step would benefit the following distributed feature encoding, since the extracted features can be less correlated by discarding the task-irrelevant correlated information across input views.
%since the task-irrelevant correlated information could be discarded.
%The extracted features can be less correlated since the task-irrelevant correlated information could be discarded.
Ideally, if each extracted feature $Z_{k}$ for $k \in \{1,\ldots,K\}$ satisfies the minimality and sufficiency conditions \cite{alemi2016deepVIB}, \cite{e22090999_ceb} as follows:
\begin{equation}
\label{equ:MNI}
\rlap{$\underbrace{\phantom{{I(X_{k} ; Z_{k}) =I(Y;X_{k})}}}_{\textbf{Minimality}}$} I(X_{k} ; Z_{k}) = \overbrace{I(Y;X_{k}) =I(Y ; Z_{k})}^{\textbf{Sufficiency}},\ k \in \{1,\ldots,K\},
%\rlap{$\underbrace{\phantom{{I(X ; Z) =I(Y;X)}}}_{\textbf{Minimality}}$} I(X ; Z) = \overbrace{I(Y;X) =I(Y ; Z)}^{\textbf{Sufficiency}}.
\end{equation}
we prove in Appendix \ref{Appendix:prove_minimal_sufficient_feature} that the minimal sufficient features meet the conditional independence assumption of the DIB theorem \cite{DVIB}.
In such cases, we will be able to obtain the rate-relevance tuples in Proposition \ref{prop:dib}.
%improve the rate-relevance tradeoff of distributed encoding.

\end{remark}

\subsection{Distributed Feature Encoding}

\label{Subsec:DIB}

% {\color{blue}
% Our distributed feature encoding step based on the DIB theorem proposed in \cite{DVIB}
% }

Once the task-relevant feature is extracted, it will be encoded for efficient transmission.
Intuitively, there is an implicit tradeoff between rate and relevance: The inference accuracy can be improved by transmitting more task-relevant information from multiple edge devices at the cost of increased communication overhead. 
By analogy with the rate-distortion tradeoff in source coding problems \cite{courtade2013multiterminal}, \cite{CEO_problem}, we resort to the DIB framework \cite{DVIB} to characterize the \emph{rate-relevance} tradeoff in edge inference.
% {\color{red} we resort to the DIB framework to characterize the}
Consider a feature encoding scheme with blocklength $N$, which maps the extracted features $Z_{k}^{1:N} = (Z_{k}^{1},\ldots,Z_{k}^{N})$ of $N$ input samples to a codeword that belongs to set $\mathcal{M}_{k}^{N}$.
The decoder at the edge server processes the codewords from devices to derive the inference result $\hat{Y}^{1:N}$.
The relevance $\Delta^{N}$ that measures the accuracy of the inference result $\hat{Y}^{1:N}$ and the rate $R_{k}^{N}$ accounting for the communication overhead of edge device $k$ are defined respectively as follows:
\begin{equation}
\label{equ:def_relevance}
    \Delta^{N} :=\frac{1}{N} I\left(Y^{1:N} ; \hat{Y}^{1:N}\right),
\end{equation}
\begin{equation}
\label{equ:def_rate}
    R_{k}^{N} \geq \frac{1}{N} \log |\mathcal{M}_{k}^{N}|, \  k \in \{1,\ldots,K\}.
\end{equation}
We endeavor to develop a set of distributed feature encoders for the edge devices and a joint decoder for the edge server that maximize the relevance {while reducing the sum of the communication costs, i.e., $R_{\text{sum}}^{N} := \sum_{k=1}^{K} R_{k}^{N}$.}
The performance of the multi-device cooperative edge inference system can be characterized by the achievable rate-relevance region, as specified in the following definition.

\begin{definition}
\label{Def:achievable_tuples}
A tuple $(\Delta, R_{\text{sum}})$ is said to be achievable if there exists a blocklength-$N$ feature encoding scheme that satisfies the following inequalities:
\begin{equation}
\label{equ:achievable_tuples}
\Delta \leq \Delta^{N} \quad \textup{and} \quad R_{\textup{sum}} \geq R_{\textup{sum}}^{N}.
\end{equation}
The {rate-relevance region} is given by the closure of all the achievable $(\Delta, R_{\text{sum}})$ tuples.
\end{definition}
The following proposition establishes a single-letter characterization of the optimal rate-relevance tuples in multi-device cooperative edge inference systems.

%The authors of \cite{DVIB} establish a single-letter characterization of the optimal rate-relevance tuples in multi-device cooperative edge inference systems. 
\begin{proposition} (Distributed Information Bottleneck \cite{DVIB})
\label{prop:dib}
Suppose the extracted features $Z_{k}$ for $k \in \{1,\ldots,K\}$ are conditionally independent given the target variable $Y$. Each $(\Delta_{\beta},R_{\beta})$ with $\beta \geq 0$ is an optimal rate-relevance tuple, i.e., there exists no relevance $\Delta \geq \Delta_{\beta}$ given the sum rate constraint $R_{\textup{sum}} = R_{\beta}$, where
%where
%i.e., there exists no relevance $\Delta \geq \Delta_{\beta}$ given the sum rate constraint $R_{\textup{sum}} = R_{\beta}$, where
\begin{equation}
    \Delta_{\beta} =  I(Y;U_{1:K}^{*}),\ R_{\beta} =  \Delta_{\beta} + \sum_{k=1}^{K} [I(Z_{k};U_{k}^{*}) - I(Y;U_{k}^{*})], \label{equ:def_rate_relevance}
\end{equation}
% \begin{align}
%     \Delta_{\beta} = & I(Y;U_{1:K}^{*}),    \label{equ:def_relevance} \\
%     R_{\beta} = & \Delta_{\beta} + \sum_{k=1}^{K} [I(Z_{k};U_{k}^{*}) - I(Y;U_{k}^{*})], \label{equ:def_rate}
% \end{align}
and the encoded features $U_{1:K}^{*}$ are obtained by minimizing the following distributed information bottleneck (DIB) objective:
\begin{equation}
\label{equ:dib}
\begin{aligned}
    \min_{\{p(\bm{u}_{k}|\bm{z}_{k})\}_{k=1}^{K}} \mathcal{L}_{\mathrm{DIB}}(\beta):= & H\left(Y |U_{1:K}\right)  + \beta \sum_{k=1}^{K}\left[H\left(Y | U_{k}\right) \right. \\
    & \left. +I\left(Z_{k} ; U_{k}\right) \right].
\end{aligned}
\end{equation}
\begin{proof}
 The proof is available in Section 7 of \cite{DVIB}.
\end{proof}

\end{proposition}

\begin{remark}
%In general the minimal sufficient conditions in (\ref{equ:MNI}) may not be satisfied \cite{wu2020learnability_ib_learnable}, and the extracted features $Z_{k}$ for $k \in \{1,\ldots,K\}$ may not be conditionally independent.
The optimality of the DIB framework relies on the conditional independence assumption, which can be satisfied when the extracted features are minimal and sufficient.
%the minimal sufficient features can be obtained 
%The optimality of the DIB framework relies on the minimal sufficient features than maintain the conditional inpendence.
Nevertheless, the minimality and sufficiency conditions in (\ref{equ:MNI}) may not be satisfied in general cases for two main reasons.
First, as discussed in \cite{DPI1_6875389}, there is a maximum value $m(X_{t},Y) \leq 1$ for dataset $(X_{t},Y)$ such that $m{(X_{t} ; Y)} \geq  \frac{I(Z_{t} ; Y)}{I(Z_{t} ; X_{t})}$ for any representation $Z_{t}$, where the Markov chain $Z_{t} \leftrightarrow X_{t} \leftrightarrow Y$ holds and the mutual information $I(Z_{t} ; X_{t}) > 0$.
% \begin{equation}
%     m{(X_{t} ; Y)}:=\sup_{Z_{t}:I(Z_{t} ; X_{t})>0} \frac{I(Z_{t} ; Y)}{I(Z_{t} ; X)}.
% \end{equation}
%This implies that we cannot find the minimal sufficient feature $Z_{t}$ 
Thus, for datasets with $m{(X_{t} ; Y)} < 1$, we cannot obtain the minimal sufficient feature $Z_{t}$.
%, the minimal sufficient feature $Z_{t}$ does not exist.
% for some datasets $(X_{t},Y)$, since there is a maximum value $m(X_{t},Y)$ such that
% \begin{equation}
%     1 \leq m_(X ; Y):=\sup _{U: U-X-Y, I(U ; X)>0} \frac{I(U ; Y)}{I(U ; X)}
% \end{equation}
Besides, the parameter $\gamma$ in the IB objective function (\ref{equ:IB_extract_final_obj}) controls the rate-relevance tradeoff and influences the solution.
Therefore, even if the minimal sufficient feature does exist, we cannot guarantee that the solution obtained by minimizing (\ref{equ:IB_extract_final_obj}) satisfies the minimality and sufficiency conditions.
%Besides, even if the minimal sufficient feature does exist, we cannot guarantee that the solution given by IB satisfies the minimality and sufficiency conditions, since the parameter $\gamma$ in the IB objective function that controls the rate-relevance tradeoff influences the solution.
%Without the perfect knowledge of the data distribution, it is challenging to select the $\gamma$ for the IB objective function in (\ref{equ:IB_extract_final_obj}).
% Selecting the appropriate $\gamma$ to make sure that
% the parameter $\gamma$ controls the rate-relevance tradeoff.
% Selecting the optimal without the perfect knowledge of .. is challenging
% complexity constraint on $I(Z_{t};X_{t})$ determined by the parameter $\beta$ may not equal to $I(X_{t};Y)$
% since the parameter $\beta$, which is associated with the complexity constraint on $I(Z_{t};X)$, cannot be appropriately selected without the perfect knowledge of the data distribution.
%should be carefully selected.
%Second, the  extracted by IB does NOT necessarily satisfy the required minimality and sufficiency conditions.
% In general one cannot guarantee that the solution given by IB is a minimal sufficient statistic – it is so only if the maximizer achieves the inequality with equality and the mutual information about Y at that point is maximum (i.e., equal I(Xk;Y).
% \cite{wu2020learnability_ib_learnable}, and the extracted features $Z_{k}$ for $k \in \{1,\ldots,K\}$ may not be conditionally independent.
But our empirical results in Appendix \ref{Appendix:ablation} demonstrate that adding an extra feature extraction step improves the performance of distributed coding. 
This can be explained by the effectiveness of feature extraction, which discards the task-irrelevant information and allows the extracted features to be less correlated compared with the raw input views $X_{k}$ for $k \in \{1,\ldots,K\}$.
\end{remark}

Proposition \ref{prop:dib} shows a distributed feature encoding scheme by minimizing the DIB objective in (\ref{equ:dib}), where the parameter $\beta$ balances the communication overhead and the inference performance.
However, obtaining the achievable rate-relevance tuples $(\Delta_{\beta},R_{\beta})$ may need a too large blocklength $N$ according to the definition in (\ref{equ:achievable_tuples}).
%However, achieving the optimal rate-relevance tuples may need a too large blocklength according to the definition in (\ref{equ:achievable_tuples}).
In other words, the edge inference system needs to accumulate a significant amount of input samples before performing distributed feature encoding, which shall postpone the inference process.
To enable instantaneous edge inference for each input sample, we reformulate the DIB framework to a distributed deterministic information bottleneck (DDIB) objective in Section \ref{Sec:Variational_Distributed_Coding}, which replaces mutual information terms $I(Z_{k};U_{k})$ with the representational costs $R_{\textup{bit}}(U_{k})$ (i.e., the number of bits) of the encoded features.
Besides, we adopt variational approximations to derive an upper bound for the conditional entropy terms $H(Y|U_{1:K})$ and $\{H(Y|U_{k})\}_{k=1}^{K}$ for more tractable optimization.

\section{Variational Approximation for Feature Extraction}
\label{Sec:Variational_feature_extraction}

In this section, we leverage a variational information bottleneck (VIB) framework to reformulate the IB objective derived in Section \ref{Subsec:IB} to facilitate tractable optimization of feature extraction.

\subsection{Variational Information Bottleneck}

Recall the IB objective in (\ref{equ:IB_extract_final_obj}) with the form of $\mathcal{L}_{\mathrm{IB},k}(\gamma) := H(Y|Z_{k}) + \gamma I(X_{k};Z_{k})$ for device $k$.
Let $p_{\bm{\theta}_{k}}(\bm{z}_{k}|\bm{x}_{k})$ denote the conditional distribution parameterized by $\bm{\theta}_{k}$ that represents the feature extraction process.
Given the joint distribution $p(\bm{x}_{k},\bm{y})$ and the Markov chain $ Y\leftrightarrow X_{k}\leftrightarrow Z_{k}$ in (\ref{equ:probability_graphical_model}), the marginal distribution $p(\bm{z}_{k})$ and the conditional distribution $p(\bm{y}|\bm{z})$ are fully determined by the conditional distribution $p_{\bm{\theta}_{k}}(\bm{z}_{k}|\bm{x}_{k})$ as follows:
\label{equ:intractable}
\begin{align} 
    p(\bm{z}_{k}) &=\int p(\bm{x}_{k}) p_{\bm{\theta}_{k}}(\bm{z}_{k} | \bm{x}_{k}) d \bm{x}_{k}, \\ 
    p(\bm{y} | \bm{z}_{k}) &= \frac{\int p(\bm{x}_{k}, \bm{y}) p_{\bm{\theta}_{k}}(\bm{z}_{k} | \bm{x}_{k}) d \bm{x}_{k} }{p(\bm{z}_{k})} . 
\end{align}
However, the distributions $p(\bm{z}_{k})$ and $p(\bm{y} | \bm{z}_{k})$ are generally intractable {due to the high dimensionality}.
Similar to \cite{shao2021learning}, we adopt tools from variational inference to approximate these intractable distributions.
{The idea of the variational approximation is to posit a family of distributions and find a member of that family which is close to the target distribution \cite{blei2017variational_survey}.}
Specifically, we introduce $r_{k}(\bm{z}_{k})$ and $p_{\bm{\varphi}_{k}}(\bm{y}|\bm{z}_{k})$ as the variational distributions to approximate $p(\bm{z}_{k})$ and $p(\bm{y} | \bm{z}_{k})$, respectively.
The family of variational conditional distributions is parameterized by $\bm{\varphi}_{k}$.
With the aid of these approximations, the VIB objective is as follows:
% {\color{blue}
% Similar to \cite{shao2021learning}, we adopt variational approximations \cite{blei2017variational_survey} to derive a tractable upper bound for the IB objective, namely, the variational information bottleneck (VIB) objective. 
% We replace the distributions $p(\bm{z}_{k})$ and $p(\bm{y} | \bm{z}_{k})$ by the variational distributions $r_{k}(\bm{z}_{k})$ and $p_{\bm{\varphi}_{k}}(\bm{y}|\bm{z}_{k})$, respectively.
% The distribution $p_{\bm{\varphi}_{k}}(\bm{y}|\bm{z}_{k})$ is parameterized by $\bm{\varphi}_{k}$.
% With the aid of these approximations, the VIB objective is as follows:
% }
\begin{equation}
\label{equ:vib}
\begin{aligned}
    \mathcal{L}_{\mathrm{VIB},k}(\gamma;\bm{\theta}_{k},\bm{\varphi}_{k}) = & \mathbf{E}_{p(\bm{x}_{k},\bm{y})}\left\{ \mathbf{E}_{p_{\bm{\theta}_{k}}(\bm{z}_{k}|\bm{x}_{k})} \left[ -\log p_{\bm{\varphi}_{k}}(\bm{y}|\bm{z}_{k}) \right] \right. \\
    & +  \gamma D_{\mathrm{KL}}(p_{\bm{\theta}_{k}}(\bm{z}_{k}|\bm{x}_{k})||r_{k}(\bm{z}_{k}))  \Big\},
\end{aligned}
\end{equation}
for which, the detailed derivation is available in Appendix \ref{Appendix:vib}.

\subsection{DNN Parameterization}

To take advantage of deep learning, we formulate the distributions $p_{\bm{\theta}_{k}}(\bm{z}_{k}|\bm{x}_{k})$ and $p_{\bm{\varphi}_{k}}(\bm{y}|\bm{z}_{k})$ in terms of DNNs with parameters $\bm{\theta}_{k}$, $\bm{\varphi}_{k}$.
%representing the DNN parameters.
A common appoarch to parameterize the condition distribution $p_{\bm{\theta}_{k}}(\bm{z}_{k}|\bm{x}_{k})$ is adopting the multivariate Gaussian distribution \cite{alemi2016deepVIB}, i.e., $p_{\bm{\theta}_{k}}(\bm{z}_{k}|\bm{x}_{k}) = \mathcal{N}(\bm{z}_{k}|\bm{\mu}_{k},\bm{\Sigma_{k}})$, where the mean vector $\bm{\mu}_{k} = (\mu_{k,1},\ldots,\mu_{k,d}) \in \mathbb{R}^{d} $ is determined by the DNN-based function $ \bm{\mu}_{k}(\bm{x}_{k};\bm{\theta}_{k})$ parameterized by $\bm{\theta}_{k}$, and the covariance matrix $\bm{\sum}_{k}$ is represented by a diagonal matrix $\diagg\{\sigma_{k,1}^{2},\ldots,\sigma_{k,d}^{2}\}$ with $\bm{\sigma}_{k} = (\sigma_{k,1},\ldots,\sigma_{k,d}) \in \mathbb{R}^{d}$ determined by the DNN-based function $\bm{\sigma}_{k}{(\bm{x}_{k};\bm{\theta}_{k})}$.
Besides, we treat the variational marginal distribution $r_{k}(\bm{z}_{k})$ as a centered isotropic Gaussian distribution $\mathcal{N}(\bm{z}_{k}|\bm{0},\bm{I})$.
%the variational marginal distribution $r_{k}(\bm{z}_{k})$ is approximated as a centered isotropic Gaussian distribution $\mathcal{N}(\bm{z}_{k}|\bm{0},\bm{I})$.
As a result, we simplify the KL-divergence term in (\ref{equ:vib}) as follows:
\begin{equation}
\label{equ:variational_KL}
\begin{aligned}
& D_{\mathrm{KL}}(p_{\bm{\theta}_{k}}(\bm{z}_{k}|\bm{x}_{k})||r_{k}(\bm{z}_{k})) \\
& = \sum_{i=1}^{d} \left\{ \frac{\mu_{k,i}^{2}+\sigma_{k,i}^{2}-1}{2} - \log{\sigma_{k,i}} \right\}.
\end{aligned}
\end{equation}

To optimize the negative log-likelihood term in (\ref{equ:vib}), we adopt the reparameterization trick \cite{kingma2013autovae} to sample $\bm{z}_{k}$ from the learned distribution $p_{\bm{\theta}_{k}}(\bm{z}_{k}|\bm{x}_{k})$, where $\bm{z}_{k} = \bm{\mu}_{k} + \bm{\sigma}_{k} \odot \bm{\epsilon_{k}}$, and $\bm{\epsilon_{k}}$ is sampled from $\mathcal{N}(\bm{0},\bm{I})$.
In this case, the Monte Carlo estimate of the negative log-likelihood term is differentiable with respect to $\bm{\theta}_{k}$.
The inference result is obtained based on the extracted feature $\bm{z}_{k}$ by the function $\bm{\hat{y}}(\bm{z}_{k};\bm{\varphi}_{k})$ with parameters $\bm{\varphi}_{k}$.
We formulate the variational conditional distribution as $p_{\bm{\varphi}_{k}}(\bm{y}|\bm{z}_{k}) \propto \exp{(-\ell(\bm{y}_{k},\bm{\hat{y}}(\bm{z}_{k};\bm{\varphi}_{k})))} $, where $\ell(\cdot,\cdot)$ denotes the loss function to measure the discrepancy between $\bm{y}$ and $\bm{\hat{y}_{k}}$.
By applying the Monte Carlo sampling method, we obtain an unbiased estimate of the gradient and hence optimize the objective function in (\ref{equ:vib}). In particular, given a minibatch of data $\{(\bm{x}_{k}^{m},\bm{y}^{m})\}_{m=1}^{M}$ at device $k$, we have the following empirical estimation of the VIB objective:
\begin{equation}
\label{equ:VIB_minibatch_optimization}
\begin{aligned}
\mathcal{L}_{\mathrm{VIB},k}(\gamma;\bm{\theta}_{k},\bm{\varphi}_{k}) \simeq & \frac{1}{M}\sum_{m=1}^{M}\left\{ -\log p_{\bm{\varphi}_{k}}(\bm{y}^{m}|\bm{z}_{k}^{m}) \right. \\
& \left. + \gamma D_{\mathrm{KL}}(p_{\bm{\theta}_{k}}(\bm{z}_{k}^{m}|\bm{x}_{k}^{m})||r_{k}(\bm{z}_{k}^{m})) \right\}.
\end{aligned}
\end{equation}
The training procedures for task-relevant feature extraction are summarized in Algorithm 1.

\begin{algorithm}[t]
%\small
\caption{Training Procedures for the Task-Relevant Feature Extraction at Device $k$}
\begin{algorithmic}[1]
\Require Training dataset $\mathcal{D}$, batch size $M$, initialized parameters $\bm{\theta}_{k}$, $\bm{\varphi}_{k}$.
\Ensure The optimized parameters $\bm{\theta}_{k}$ and $\bm{\varphi}_{k}$
\Repeat
\State Randomly select a minibatch $\left\{\left(\bm{x}_{k}^{m}, \bm{y}^{m}\right)\right\}_{m=1}^{M}$ from $\mathcal{D}$.
\State Compute the mean vector $\{\bm{\mu}_{k}^{m}\}_{m=1}^{M}$ and the standard 

deviation vector $\{\bm{\sigma}_{k}^{m}\}_{m=1}^{M}$.
\While{$m=1$ to $M$}
\State Sample $\bm{\epsilon_{k}^{m}} \sim \mathcal{N}(\bm{0},\bm{I})$.
\State Compute $\bm{z}_{k}^{m} = \bm{\mu}_{k}^{m} + \bm{\sigma}_{k}^{m} \odot \bm{\epsilon_{k}^{m}}.$
\EndWhile
\State Compute the loss $\mathcal{L}_{\mathrm{VIB},k}(\gamma;\bm{\theta}_{k},\bm{\varphi}_{k})$ based on (\ref{equ:VIB_minibatch_optimization}).
\State Update parameters $\bm{\theta}_{k},\bm{\varphi}_{k}$ through backpropagation.
\Until{Convergence of parameters $\bm{\theta}_{k}$ and $\bm{\varphi}_{k}$}
\end{algorithmic}
\label{algorithm:VIB}
\end{algorithm}

\section{Variational Distributed Feature Encoding}
\label{Sec:Variational_Distributed_Coding}
In this section, we develop a distributed feature encoding scheme based on the DIB objective in (\ref{equ:dib}) to achieve efficient task-oriented communication.
We first reformulate the DIB objective to a \emph{distributed deterministic information bottleneck} (DDIB) objective that explicitly takes into account the communication overhead of each encoded feature.
Besides, variational approximations are invoked again to make the optimization of the DDIB objective tractable.

\subsection{Distributed Deterministic Information Bottleneck Reformulation}
\label{Section:VDDIB}
The DIB objective in (\ref{equ:dib}) measures the data rate by the mutual information terms $I(Z_{k},U_{k})$ for $k \in \{1,\ldots,K\}$.
Although the mutual information has broad applications in representation learning \cite{DVIB,goldfeld2020information_B_application}, directly controlling the amount of resources required to represent each $U_{k}$ (i.e., the number of bits) is {more relevant in communication systems}.
To address the communication cost, we select a deterministic encoding function that maps each task-relevant feature $Z_{k}$ to a bitstream $U_{k}$, and adopt the number of bits, denoted as $R_{\text{bit}}(U_{k})$, to measure the communication overhead, which results in the following DDIB objective:
\begin{equation}
\label{equ:ddib}
    \mathcal{L}_{\mathrm{DDIB}}(\beta):=H\left(Y |U_{1:K}\right) + \beta \sum_{k=1}^{K}\left[H\left(Y | U_{k}\right) +R_{\text{bit}}\left( U_{k}\right) \right].
\end{equation}
In particular, the feature encoder in each edge device first performs the encoding transformation $\bm{f}_{\bm{\phi}_{k}}$, followed by the uniform quantizer $\bm{\mathcal{Q}}$ that discretizes each element in $\bm{f}_{\bm{\phi}_{k}}(\bm{z}_{k}) \in \mathbb{R}^{d_{k}}$ into $n_{k}$ bits.
The encoded feature is given by $\bm{u}_{k} := \bm{\mathcal{Q}}(\bm{f}_{\bm{\phi}_{k}}(\bm{z}_{k}))$, and the gradients in the training process is approximated by the straight-through estimator (STE) \cite{bengio2013estimating_STE}.
%, i.e., the encoded feature is given by $\bm{u}_{k} = \bm{\mathcal{Q}}(\bm{f}_{\bm{\phi}_{k}}(\bm{z}_{k}))$.
%We select each $\bm{\mathcal{Q}}$ as a uniform quantizer that discretizes each element in $\bm{f}_{\bm{\phi}_{k}}(\bm{z}_{k}) \in \mathbb{R}^{d_{k}}$ into $n_{k}$ bits, and leverages the straight-through estimator (STE) \cite{bengio2013estimating_STE} to approximate the gradients in the training process. 
The amount of information needed to be transmitted from the device $k$ to the edge server is $R_{\rm{bit}}(\bm{u}_{k}) = n_k d_k$ bits\footnote{Note that the quantization scheme is orthogonal to the proposed DDIB method. We select the uniform quantization in this paper, since it is easier to approximate the gradients and calculate the bit-length.}.
%In other words, $R_{\rm{bit}}(\bm{u}_{k}) = n_k d_k$ bits need to be transmitted in order to transfer the encoded feature $\bm{u}_k$ from device $k$ to the edge server\footnote{Note that the quantization scheme is orthogonal to the proposed DDIB method. We select the uniform quantization in this paper, since it is easier to approximate the gradients and calculate the bit-length.}.

\subsection{Variational Distributed Deterministic Information Bottleneck}
The conditional distributions $\{p(\bm{y}|\bm{u}_{k})\}_{k=1}^{K}$ and $p(\bm{y}|\bm{u}_{1:K})$ are fully determined given deterministic mappings from $Z_k$ to $U_k$, $k \in \{ 1,\cdots, K \}$. 
However, calculating the conditional entropy terms $H(Y|U_{1:K})$ and $H(Y|U_{k})$ in (\ref{equ:ddib}) is generally intractable due to the high-dimensional integrals.
Following Section \ref{Sec:Variational_feature_extraction}, we adopt variational distributions $p_{\bm{\psi}_{0}}(\bm{y}|\bm{u}_{1:K})$ and $\{p_{\bm{\psi}_{k}}(\bm{y}|\bm{u}_{k})\}_{k=1}^{K}$ to replace the terms $p(\bm{y}|\bm{u}_{1:K})$ and $\{p(\bm{y}|\bm{u}_{k})\}_{k=1}^{K}$.
Denoting $\bm{\psi} := \{\bm{\psi_{0}},\{\bm{\psi}_{k}\}_{k=1}^{K}\}$ and $\bm{\phi} := \{\bm{\phi}_{k}\}_{k=1}^{K}$, we define a \emph{variational distributed deterministic information bottleneck} (VDDIB) objective as follows:
\begin{equation}
\label{equ:vddib}
\begin{aligned}
    \mathcal{L}_{\mathrm{VDDIB}} (\beta;\bm{\phi},\bm{\psi}) :=  \mathbf{E} \Bigg\{  - \log p_{\bm{\psi_{0}}}(\bm{y}|\bm{u}_{1:K})   \\
     + \beta \Bigg\{  \sum_{k=1}^{K}   - \log{p_{\bm{\psi}_{k}}(\bm{y}|\bm{u}_{k})}  + \sum_{k=1}^{K} R_{\text{bit}}(\bm{u}_{k}) \Bigg\} \Bigg\} ,
     %\text{with} \quad \bm{u}_{k} = \bm{\mathcal{Q}}(\bm{f}_{\bm{\phi}_{k}}(\bm{z}_{k})), \\
     % \text{with} \quad p_{\bm{\theta}}(\bm{z}_{1:K},\bm{y})= \int p(\bm{x}_{1:K},\bm{y}) \prod_{k=1}^{K} p_{\bm{\theta}_{k}}(\bm{z}_{k}|\bm{x}_{k})d\bm{x}_{1:K},
\end{aligned}
\end{equation}
where the expectation is taken over the joint distribution of $Y$ and $U_{1:K}$.
% where the parameters $\{\bm{\theta}_{k}\}_{k=1}^{K}$ are obtained by the task-relevant feature extraction learned from Algorithm 1.
The VDDIB objective is an upper bound of the DDIB objective as shown in Appendix \ref{Appendix:VDDIB}.
The variational distributions are formulated as follows:
\begin{equation}
p_{\bm{\psi}_{0}}(\bm{y}|\bm{u}_{1:K})\propto \exp(-\ell(\bm{y},\bm{\hat{y}}(\bm{u}_{1:K};\bm{\psi}_{0}))),
\end{equation}
\begin{equation}
    p_{\bm{\psi}_{k}}(\bm{y}|\bm{u}_{k}) \propto \exp(-\ell(\bm{y},\bm{\hat{y}}(\bm{u}_{k};\bm{\psi}_{k}))), k \in \{1,\ldots,K\},
\end{equation}
where the functions $\bm{\hat{y}}(\bm{u}_{1:K};\bm{\psi}_{0})$ and $\{\bm{\hat{y}}(\bm{u}_{k};\bm{\psi}_{k})\}_{k=1}^{K}$ can be interpreted as the predictors that use the received features to predict the target variable.
In the training process, by leveraging the Monte Carlo sampling method to select a minibatch $\{(\bm{z}_{1:K}^{m},\bm{y}^{m})\}_{m=1}^{M}$, we have the following expression to estimate the VDDIB objective function:
% In the training process, by leveraging the Monte Carlo sampling method as in (\ref{equ:VIB_minibatch_optimization}) and selecting a minibatch $\{(\bm{z}_{1:K}^{m},\bm{y}^{m})\}_{m=1}^{M}$ from the joint distribution $p_{\bm{\theta}}(\bm{z}_{1:K},\bm{y})$, we have the following expression to estimate the VDDIB objective function:
\begin{equation}
\label{equ:vddib_Monto_Carlo}
\begin{aligned}
\mathcal{L}_{\mathrm{VDDIB}} (\beta;\bm{\phi},\bm{\psi}) \simeq \frac{1}{M} \sum_{m=1}^{M} \Bigg\{  - \log p_{\bm{\psi_{0}}}(\bm{y}^{m}|\bm{u}_{1:K}^{m}) . \\
+ \left. \beta \sum_{k=1}^{K}  \left\{ - \log{p_{\bm{\psi}_{k}}(\bm{y}^{m}|\bm{u}_{k}^{m})} + R_{\text{bit}}(\bm{u}_{k}^{m}) \right\} \right\}.
\end{aligned}
\end{equation}
The optimization procedures for the VDDIB objective are summarized in Algorithm 2.

\begin{algorithm}[t]
\small
\caption{Training Procedures for the VDDIB Method}
\begin{algorithmic}[1]
\Require Training dataset $\mathcal{D}$, batch size $M$, optimized parameters $\{\bm{\theta}_{k}\}_{k=1}^{K}$, and initialized parameters $\bm{\phi}$, $\bm{\psi}$.
\Ensure The optimized parameters $\bm{\phi}$ and $\bm{\psi}$.
\Repeat
\State Randomly select a minibatch $\left\{\left(\bm{x}_{1:K}^{m}, \bm{y}^{m}\right)\right\}_{m=1}^{M}$ from $\mathcal{D}$.
\State Extract the task-relevant features $\left\{\bm{z}_{1:K}^{m}\right\}_{m=1}^{M}$ based on the 

learned distributions $\{p_{\bm{\theta}_{k}}(\bm{z}_{k}|\bm{x}_{k})\}_{k=1}^{K}$.
\State Compute the loss $\mathcal{L}_{\mathrm{VDDIB}}(\beta;\bm{\phi},\bm{\psi})$ based on (\ref{equ:vddib_Monto_Carlo}).
\State Update parameters $\bm{\phi},\bm{\psi}$ through backpropagation.
%\EndWhile
\Until{Convergence of parameters $\bm{\phi}$ and $\bm{\psi}$.}
%\State 
\end{algorithmic}
\label{algorithm:VDDIB}
\end{algorithm}

\begin{remark}
\label{remark:VDDIB}
Given the random variables $U_{1:K}$, the VDDIB objective is an upper bound of the DIB objective as shown in the following expression, owing to $I(Z_{k},U_{k}) \leq H(U_{k}) \leq R_{\text{bit}}(U_{k})$ and the variational approximations:
\begin{equation}
\label{equ:DDIB_VDDIB_inequality}
\mathcal{L}_{\mathrm{DIB}}(\beta) \leq \mathcal{L}_{\mathrm{DDIB}}(\beta) \leq
 \mathcal{L}_{\mathrm{VDDIB}}(\beta;\bm{\phi},\bm{\psi}).
\end{equation}
It is important to note that the distributed feature encoding scheme obtained by minimizing the VDDIB objective does not guarantee to achieve the optimal rate-relevance tuples in Proposition \ref{prop:dib} due to the approximations in the IB and DIB optimizations.
Nevertheless, the empirical results in Section \ref{Sec_shape_reco} evidence that the proposed method notably outperforms the existing communication strategies for multi-device cooperative edge inference.
\end{remark}

\section{Distributed Feature Encoding with Selective Retransmission}
\label{Sec:Selective_Transmission}
As highlighted in Remark \ref{remark:VDDIB}, minimizing the VDDIB objective may not result in the optimal rate-relevance tuples.
To further reduce the communication overhead, we need more effective approaches to identify the redundancy in the encoded features of multiple edge devices.
%For this purpose, we propose a VDDIB-SR method that introduces a \emph{selective retransmission} (SR) mechanism to the VDDIB framework, where the edge server selectively activates the edge devices to retransmit their encoded features based on the informativeness of the received features.
For this purpose, we propose a VDDIB-SR method that introduces a \emph{selective retransmission} (SR) mechanism to the VDDIB framework, which performs multiple sequential communication rounds \cite{kurka2019successive_refinement} between the edge devices and the edge server.
Particularly, the edge server selectively activates the edge devices to retransmit their encoded features based on the informativeness of the received features.

\subsection{Selective Retransmission Mechanism}

%\footnote{In the following, we also use $\{\bm{u}_{k,t}\}_{t=1}^{T}$ to denote $[\bm{u}_{k,1},\ldots,\bm{u}_{k,T}]$.}

In this framework, each device is allowed to transmit the encoded feature vector $\bm{u}_{k} = [\bm{u}_{k,1},\ldots,\bm{u}_{k,T}]$ with a maximum number of $T$ attempts (i.e., $T-1$ retransmissions).
The deterministic mapping from $Z_{k}$ to $U_{k}$ is defined as $ \bm{u}_{k} = \bm{\mathcal{Q}}(\bm{f}_{\bm{\widetilde{\phi}}_{k}}(\bm{z}_{k}))$.
The feature encoding transformation $\bm{f}_{\bm{\widetilde{\phi}}_{k}}$ is a DNN-based function parameterized by $\bm{\widetilde{\phi}}_{k}$.
The parameters $\{\bm{\widetilde{\phi}}_{k}\}_{k=1}^{K}$ are defined as $\bm{\widetilde{\phi}}$ for simplicity.
The edge server has $T$ independent predictors, where the $t$-th predictor outputs the inference result based on the received features after the $t$-th transmission.
Once the received features are sufficient to output a confident result, the remaining retransmission attempts can be saved, {which is favorable for latency reduction.}
To unleash the full benefits of the selective retransmission mechanism, it is critical to design a \emph{stopping policy} to decide when to terminate the retransmission process.
Besides, as the encoded features of multiple edge devices may be redundant for edge inference, it leaves room for further communication overhead reduction by scheduling the edge devices with the most informative features for retransmissions.
Thus, an \emph{attention module} is proposed that activates the edge devices with the most informative features to retransmit.
The received features from the activated devices in the $t$-th transmission are defined as $\bm{\tilde{u}}_{1:K,t}$, where $\bm{\tilde{u}}_{k,t} =\bm{{u}}_{k,t}$ for the activated devices, and $\bm{\tilde{u}}_{k,t} =\bm{0}$ otherwise.

\begin{figure*}[t]
    \centering
    \subfloat[The stopping policy that determines when to terminate the retransmission process.]{
    \label{fig:SR-a}
    \centering
    \includegraphics[width=0.8\linewidth]{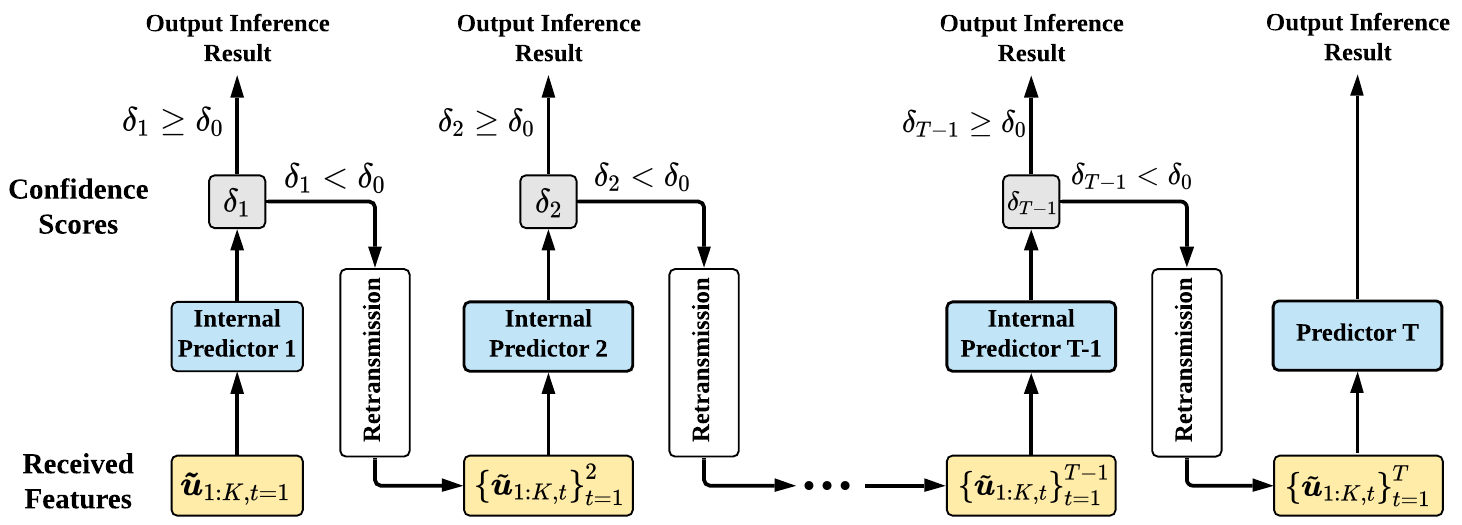}
    }
    \vfill
    \subfloat[The attention module that determines which devices to retransmit.]{
    \centering
    \label{fig:SR-b}
    \includegraphics[width=0.705\linewidth]{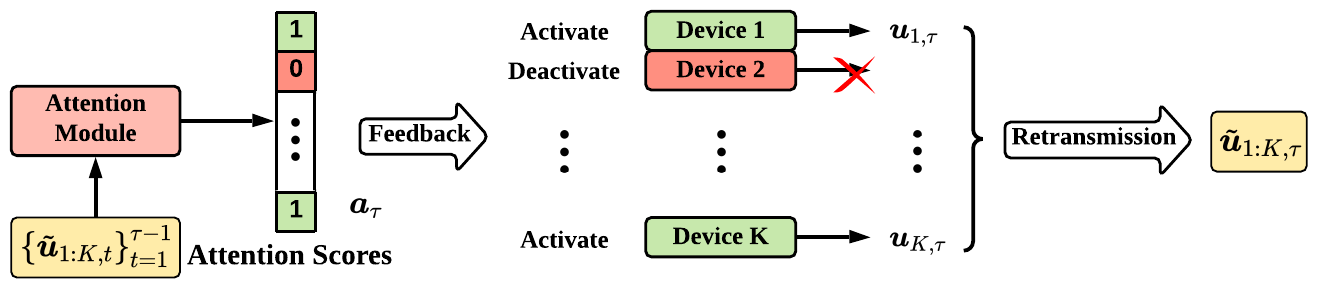}
    }
    \caption{The selective retransmission mechanism with (a) a stopping policy and (b) an attention module.}
    \label{fig:SR-framework}
\end{figure*}

\subsubsection{Stopping Policy}
Inspired by cascade inference \cite{chen2020learning_to_stop}, \cite{enomoto2021learning_to_cascade}, we develop a confidence-based stopping policy at the edge server that outputs the confidence scores $\delta_{\tau}$, $\tau \in \{1,\ldots,T-1\}$, of the inference results after each transmission attempt.
Its structure is depicted in Fig. \ref{fig:SR-a}, which consists of $T-1$ internal predictors, and the confidence scores determine when to terminate the retransmission process.
Specifically, the $\tau$-th internal predictor outputs the inference result based on the received features $\{\bm{\tilde{u}}_{1:K,t}\}_{t=1}^{\tau}$ defined as follows:
\begin{equation}
\label{equ:cascade_predictor}
    \textbf{Internal Predictors:}  \ \bm{\hat{y}}\left(\{\bm{\tilde{u}}_{1:K,t}\}_{t=1}^{\tau};\bm{\widetilde{\psi}}_{\tau}\right), \tau \in \{1,\ldots,T\},
\end{equation}
where the notation $\bm{\widetilde{\psi}}_{\tau}$ represents the DNN parameters that form the $\tau$-th internal predictor, and we define $\bm{\widetilde{\psi}}:=\{\bm{\widetilde{\psi}_{t}}\}_{t=1}^{T}$ for notational simplicity. The corresponding variational distributions are as follows:
\begin{equation}
\label{equ:VDDIB-SR_Variational_distributions}
\begin{aligned}
p_{\bm{\widetilde{\psi}}_{\tau}}(\bm{y}|\{\bm{\tilde{u}}_{1:K,t}\}_{t=1}^{\tau}) & \propto \exp{\left(-\ell(\bm{y},\bm{\hat{y}}\left(\{\bm{\tilde{u}}_{1:K,t}\}_{t=1}^{\tau};\bm{\widetilde{\psi}}_{\tau}\right)\right)}, \\
 \quad \text{with} \quad  \tau & \in \{1,\ldots,T\}.
\end{aligned}
\end{equation}

A threshold value $\delta_{0}$ is selected to decide when to terminate the retransmission process:
If a confidence score $\delta_{\tau}$ is above $\delta_{0}$, the intermediate inference result is likely to be correct, and the inference process is terminated with that result;
Otherwise, more features from the edge devices are preferred by the edge server to improve the inference performance.
In other words, the threshold value $\delta_{0}$ allows us to flexibly control the tradeoff between the inference performance and the communication overhead.
Particularly, we use the {maximum predicted probability} as the confidence score for the classification tasks expressed as follows:
\begin{equation}
\label{equ:confidence_score}
\begin{aligned}
 \textbf{Confidence Scores:} \ \delta_{\tau} & = \max\left(\bm{\hat{y}}\left(\{\bm{\tilde{u}}_{1:K,t}\}_{t=1}^{\tau};\bm{\widetilde{\psi}}_{\tau}\right)\right), \\
 \text{with} \quad \tau & \in \{1,\ldots,T-1\}.
\end{aligned}
\end{equation}

\subsubsection{Attention Module}
The attention module is developed to determine which devices should be activated for retransmissions.
Each attention score $a_{k,\tau} \in \{0,1\}$ for the edge device $k$ in the $\tau$-th transmission is based on the received features $\{\bm{\tilde{u}}_{1:K,t}\}_{t=1}^{\tau-1}$ determined as follows:
\begin{equation}
\label{equ:attention_score}
\begin{aligned}
    \textbf{Attention Scores:} \ a_{k,\tau} = g_{k,\tau-1}(\{\bm{\tilde{u}}_{1:K,t}\}_{t=1}^{\tau-1}), \\
     \text{with} \quad k \in \{1,\ldots,K\} \ \text{and} \ \tau \in \{2,\ldots,T\},
\end{aligned}
\end{equation}
where functions $ \{g_{k,1},\ldots,g_{k,T-1}\}_{k=1}^{K}$ are constructed by fully-connected layers.
Note that we use a hard binarization function to discretize the attention scores and select the straight-through estimator to approximate the gradients.
Particularly, we set $a_{k,1} = 1$, $k \in \{1,\ldots,K\}$, which means that all the edge devices need to transmit their features in the first transmission.
For notational simplicity, we define the attention vectors as $\bm{a_{\tau}} := \{a_{k,\tau}\}_{k=1}^{K}$, $\tau \in \{1,\ldots,T\}$.
As illustrated in Fig. \ref{fig:SR-b}, if the attention score $a_{k,\tau}$ is 1, the feature $\bm{u}_{k,\tau}$ would be transmitted from the edge device $k$ to the edge server;
Otherwise, the edge device $k$ would be deactivated in the $\tau$-th transmission.
In other words, we have $\bm{\tilde{u}}_{k,\tau} = a_{k,\tau} \bm{u}_{k,\tau}$.

\subsection{VDDIB-SR Objective}

%By adopting the stopping policy and the attention module, the VDDIB-SR objective is defined as follows:
By adopting the stopping policy and introducing the attention module, we extend the VDDIB objective in (\ref{equ:vddib}) to the VDDIB-SR objective as follows:
\begin{equation}
\label{equ:vddib-sr}
\begin{aligned}
& \mathcal{L}_{\mathrm{VDDIB-SR}}  \left(\beta,T;\bm{\widetilde{\phi}},\widetilde{\bm{\psi}},\{\bm{\psi}_{k}\}_{k=1}^{K}\right) \\
& :=  \mathbf{E} \Bigg\{ \frac{1}{T} \sum_{\tau=1}^{T}  - \log {p_{\bm{\widetilde{\psi}}_{\tau}}(\bm{y}|\{\bm{\tilde{u}}_{1:K,t}\}_{t=1}^{\tau})} \\
& \quad + \beta \Bigg \{ \sum_{k=1}^{K} - \log {p_{\bm{\psi}_{k}}(\bm{y}|\bm{u}_{k})} + \sum_{k=1}^{K}\sum_{t=1}^{T} R_{\text{bit}}(\bm{\tilde{u}}_{k,t})   \Bigg\} \Bigg\}.
\end{aligned}
\end{equation}
{In (\ref{equ:vddib-sr}), the total loss of the internal predictors is shown in the first term that takes into account all possible stopping points.
The second term in the VDDIB-SR objective acts as an auxiliary loss, which maintains the informativeness of each encoded feature and makes the training process robust against dynamic activation caused by the attention module.
Besides, each $R_{\text{bit}}(\bm{\tilde{u}}_{k,t})=a_{k,t}n_{k}d_{k}$ in the third term of the VDDIB-SR objective measures the communication overhead at the $t$-th transmission of the edge device $k$. }

By applying the Monte Carlo sampling method, we obtain an unbiased estimate of the gradient and hence minimize the objective in (\ref{equ:vddib-sr}) using backpropagation. Given a minibatch of data $\{\bm{x}_{1:K}^{m},\bm{y}^{m}\}_{m=1}^{M}$, we can extract the task-relevant features $\{\bm{z}_{1:K}^{m}\}_{m=1}^{M}$ by the learned parameters $\{\bm{\theta}_{k}\}_{k=1}^{K}$ in the VIB framework.
The Monte Carlo estimate of $\mathcal{L}_{\mathrm{VDDIB-SR}}$ is given as follows:
\begin{equation}
\label{equ:vddib-sr_monte_carlo}
\begin{aligned}
 &  \mathcal{L}_{\mathrm{VDDIB-SR}}\left(\beta,T;\bm{\widetilde{\phi}},\widetilde{\bm{\psi}},\{\bm{\psi}_{k}\}_{k=1}^{K}\right) \\
 &  \simeq   \frac{1}{M}  \left\{ \frac{1}{T} \sum_{\tau=1}^{T}  - \log {p_{\bm{\widetilde{\psi}}_{\tau}}(\bm{y}^{m}|\{\bm{\tilde{u}}_{1:K,t}^{m}\}_{t=1}^{\tau})}  \right. \\
 &  \quad + \beta \sum_{k=1}^{K} \left\{ - \log {p_{\bm{\psi}_{k}}(\bm{y}^{m}|\bm{u}_{k}^{m})} + \sum_{t=1}^{T} R_{\text{bit}}(\bm{\tilde{u}}_{k,t}^{m})  \right\} \Bigg\}.
\end{aligned}
\end{equation}
The training procedures for the VDDIB-SR method are summarized in Algorithm 3.

\begin{algorithm}[t]
\small
\caption{Training Procedures for the VDDIB-SR Method}
\begin{algorithmic}[1]
\Require Training dataset $\mathcal{D}$, batch size $M$, optimized parameters $\{\bm{\theta}_{k}\}_{k=1}^{K}$, and initialized parameters $\bm{\widetilde{\phi}}$, $\bm{\widetilde{\psi}}$, and $\{\bm{\psi}_{k}\}_{k=1}^{K}$.
\Ensure The optimized parameters $\bm{\widetilde{\phi}}$, $\bm{\widetilde{\psi}}$, and $\{\bm{\psi}_{k}\}_{k=1}^{K}$.
\Repeat
%\While{epoch $t=1$ to $T$ {\color{red} change while to repeat}}
\State Randomly select a minibatch $\left\{\left(\bm{x}_{1:K}^{m}, \bm{y}^{m}\right)\right\}_{m=1}^{M}$ from $\mathcal{D}$.
\State Extract the task-relevant features $\left\{\bm{z}_{1:K}^{m}\right\}_{m=1}^{M}$ based on the

learned distributions $\{p_{\bm{\theta}_{k}}(\bm{z}_{k}|\bm{x}_{k})\}_{k=1}^{K}$.
\While{$\tau=2$ to $T$}
\State Compute the attention scores $a_{k,\tau}$ for $k \in 1: K$.
\State Determine the features $\tilde{\bm{u}}_{k, \tau}:=a_{k, \tau} \bm{u}_{k, \tau}$ for $k \in 1: K$.
\State Obtain the {\color{black}intermediate} inference result based on (\ref{equ:cascade_predictor}).
\EndWhile
\State Compute the loss $\mathcal{L}_{\mathrm{VDDIB-SR}}\left(\beta,T;\bm{\widetilde{\phi}},\widetilde{\bm{\psi}},\{\bm{\psi}_{k}\}_{k=1}^{K}\right)$ based 

on (\ref{equ:vddib-sr_monte_carlo}).
\State Update parameters $\bm{\widetilde{\phi}}$, $\bm{\widetilde{\psi}}$, and $\{\bm{\psi}_{k}\}_{k=1}^{K}$ through 

backpropagation.
%\EndWhile
\Until{Convergence of parameters $\bm{\widetilde{\phi}}$, $\bm{\widetilde{\psi}}$, and $\{\bm{\psi}_{k}\}_{k=1}^{K}$.}
%\State 
\end{algorithmic}
\label{algorithm:VDDIB-SR}
\end{algorithm}

\begin{remark}
\label{remark:VDDIB-SR1}
%The VDDIB-SR objective is an lower bound of the VDDIB objective in (\ref{equ:vddib}), and the minimum value of $\mathcal{L}_{\mathrm{VDDIB-SR}}\left(\beta,T;\bm{\widetilde{\phi}},\widetilde{\bm{\psi}},\{\bm{\psi}_{k}\}_{k=1}^{K}\right)$ is a decreasing function of $T$.
%The VDDIB-SR objective is an lower bound of the VDDIB objective in (\ref{equ:vddib}).
%Denote the parameters in the VDDIB-SR objective $\mathcal{L}_{\mathrm{VDDIB-SR}}(\beta,T)$ as $\bm{P}_{T}$ for notational simplicity.
%Denote the parameter space of the VDDIB-SR objective as $\bm{P}_{T}$ for notational simplicity.
Denote the parameter space of the VDDIB-SR objective as $\bm{P}_{T}$.
It is straightforward that $\bm{P_{1}} \subseteq \dots \subseteq \bm{P}_{T}$, where $\bm{P_{1}}$ represents the parameters of the VDDIB objective.
Thus, it implies that the minimum of the VDDIB-SR objective is a lower bound of the VDDIB objective in (\ref{equ:vddib}), and increasing the maximum transmission attempts $T$ could achieve a better rate-relevance tradeoff.
Particularly, let $\bm{P} = \{p(\bm{u}_{1:K}|\bm{z}_{1:K})\}$ denote the distributions of the joint feature encoding scheme. 
The minimum of $\mathcal{L}_{\mathrm{DDIB}}(\beta)$ over $\bm{P}$ lower bounds $\mathcal{L}_{\mathrm{VDDIB-SR}}\left(\beta,T;\bm{\widetilde{\phi}},\widetilde{\bm{\psi}},\{\bm{\psi}_{k}\}_{k=1}^{K}\right)$ for any integer $T$.
%The minimum value of the DDIB objective $\min_{\bm{P}} \mathcal{L}_{\text{DDIB}}(\beta)$ lower bounds $\mathcal{L}_{\mathrm{VDDIB-SR}}\left(\beta,T;\bm{\widetilde{\phi}},\widetilde{\bm{\psi}},\{\bm{\psi}_{k}\}_{k=1}^{K}\right)$ for any integer $T$.
This is consistent with the classic source coding theory \cite{network_information_theory} that the performance of the joint coding schemes is better than that of the distributed coding methods.
The detailed derivation is deferred to Appendix \ref{appendix:vddib-sr}.

\end{remark}

\section{Performance Evaluation}
\label{Sec:Experiment}

In this section, we evaluate the performance of the proposed task-oriented communication methods for cooperative edge inference on two kinds of cooperative inference tasks, including multi-view image classification and multi-view object recognition.

\subsection{Experimental Setup}

\subsubsection{Datasets}
In the multi-view image classification task, we select MNIST \cite{lecun1998gradientMNIST}, CIFAR-10 \cite{cifar10/100}, and Tiny ImageNet \cite{tiny_imagenet} for experiments.
The MNIST dataset contains handwritten digit images in 10 classes, which has a training set of 60,000 examples and a test set of 10,000 examples. 
The CIFAR-10 dataset consists of 60,000 color images in 10 classes (6,000 images per class).
Particularly, 5,000 images of each class are used for training, and the remaining 1,000 images are used for testing.
Tiny ImageNet contains 110,000 images of 200 classes (550 images per class).
Each class has 500 training images and 50 validation images.
We follow the method in \cite{Neural_DSC} to generate two views for each image by splitting it vertically. 
Each view has the same size, and we assume that there are two devices with distinct views in the edge inference system.

In the multi-view object recognition task, we use Washington RGB-D (WRGBD) \cite{WRGBD} and ModelNet40 \cite{wu20153d_shapenet40} to evaluate the proposed methods.
The WRGBD dataset includes household objects organized into 51 classes.
Each object is captured by three cameras mounted at different heights, at $30^{\circ}$, $45^{\circ}$, and $60^{\circ}$ above the horizon.
We select 80\% of objects from each class for training and use the remaining objects for testing.
Thus, there are 19,390 objects in the training set and 2,729 objects in the test set.
ModelNet40 contains 12,311 computer-aided design (CAD) objects of 40 categories.
Similar to a prior study \cite{su2015multi_MVCNN}, each object is augmented by rotating 30 degrees along the gravity direction, resulting in twelve views.
We choose around 80 objects per class for training and 20 objects per class for testing.

% We use the MNIST dataset \cite{lecun1998gradientMNIST} for multi-view image classification, which consists of 28×28 handwritten digit images in 10 classes.
% Each class has 6,000 samples for training and 1,000 samples for testing.
% We vertically split each image to two views and assume that there are two devices with distinct views in the edge inference system.
% Besides, we select the ModelNet40 dataset \cite{wu20153d_shapenet40} for the multi-view object recognition task, which contains 12,311 computer-aided design (CAD) {\color{blue}{objects}} of 40 categories.
% Each object is augmented by rotating 30 degrees along the gravity direction, resulting in twelve views.
% Similar to a prior study \cite{su2015multi_MVCNN}, we choose around 80 objects per class for training and 20 objects per class for testing.

\subsubsection{Proposed methods} We investigate the performance of the proposed VDDIB and VDDIB-SR methods.
The VDDIB-SR method with $t$ transmissions is denoted as VDDIB-SR (T=t).
Particularly, the VDDIB method can be regarded as a special case of the VDDIB-SR method with $T=1$.
In the training process, we set $\gamma=10^{-4}$ when minimizing the VIB objective $\mathcal{L}_{\mathrm{VIB},k}(\gamma)$ for feature extraction.
{To achieve different rate-relevance tuples, the hyperparameter $\beta$ varies in the range of $[10^{-3},10^{-1}]$ when optimizing the VDDIB and VDDIB-SR objectives}\footnote{The code is available at https://github.com/shaojiawei07/VDDIB-SR.}.

\subsubsection{Baselines} There are only a few existing studies that are relevant to feature encoding for cooperative edge inference. We compare the proposed methods against the following five baseline methods:
\begin{itemize}
\item Server-based inference: This baseline deploys a powerful DNN on the edge server for inference and adopts the data-oriented communication scheme to transmit the multiple views. The view size is reduced by using PNG compression.
%We reduce the image size using PNG compression
%from the edge devices to the edge server.
\item NN-REG and NN-GBI \cite{hanna2020distributed}: {Both of them adopt learning-based methods to quantize the input data to bitstreams at the edge device}. The DNN-based quantizers are either trained using the quantization loss regularization (NN-REG) or with the greedy boundary insertion (GBI) algorithm (NN-GBI)\footnote{\color{black}To reduce the search space and improve the training efficiency, we consider only one possible boundary for each dimension.}.
\item eSAFS \cite{singhal2020communication}: In this scheme, each edge device extracts a feature vector as well as a small-scale importance vector using DNNs. These importance vectors are transmitted to an edge server for preprocessing, which identifies the important feature vectors that need to be transmitted by the edge devices.
\item CAFS \cite{choi2019context}: This method integrates a context-aware feature selection (CAFS) scheme into the multi-device cooperative edge inference system. {Each edge device extracts a task-relevant feature and determines its importance via entropy-based likelihood estimation}. 
The features from different devices are aggregated using a weighted mean of their likelihoods. By regulating the likelihood cut-off level, edge devices with less important features are deactivated.
\end{itemize}

\begin{table}[t]
\caption{The structures of NN$_1$ and NN$_2$ for different tasks.} 
\centering
\resizebox{0.495\textwidth}{!}{
\begin{tabular}{c|c|c|c|c|c}
\hline
    & MNIST       & CIFAR-10    & Tiny-ImageNet                                                      & WRGBD                                                              & ModelNet40  \\ \hline
NN$_1$ & 3 FC layers & VGG-19 \cite{VGG}      & PreActResNet18 \cite{PreActResNet}                                                     & \begin{tabular}[c]{@{}c@{}}2 Conv layers\\ 1 FC layer\end{tabular} & VGG-11 \cite{VGG}      \\ \hline
NN$_2$ & 3 FC layers & 2 FC layers & \begin{tabular}[c]{@{}c@{}}2 Conv layers\\ 1 FC layer\end{tabular} & 1 FC layer                                                         & 3 FC layers \\ \hline
\end{tabular}
}
\label{Table:NN1_NN2_structure}
\end{table}

\begin{table*}[]
\caption{The neural network architectures of different methods.} 
\begin{center}
%\begin{threeparttable}
\resizebox{0.99\textwidth}{!}{
\begin{tabular}{c|c|c|c|c|c|c|c}
\hline
                                         &                   
                                         \begin{tabular}[c]{@{}c@{}}\textbf{Server-based}\\ \textbf{Inference}\end{tabular}                            & \textbf{NN-REG}                                                        & \textbf{NN-GBI}                                                        & \textbf{eSAFS}                                                           & \textbf{CAFS}                                                     & \textbf{VDDIB}                                                         & \textbf{VDDIB-SR}                                                   \\ \hline
\multirow{3}{*}{\textbf{\begin{tabular}[c]{@{}c@{}}   \\ On-device \\ Network\end{tabular}}} &  \multirow{3}{*}{\textbf{\begin{tabular}[c]{@{}c@{}}   \\  \\ ---\end{tabular}}}   & \multirow{2}{*}{\begin{tabular}[c]{@{}c@{}} \\ NN$_1$ \end{tabular}}                                                 & \multirow{2}{*}{\begin{tabular}[c]{@{}c@{}} \\ NN$_1$ \end{tabular}}                                                    & \multirow{2}{*}{\begin{tabular}[c]{@{}c@{}} \\ NN$_1$ \end{tabular}}                                                      & NN$_1$                                                               & \multirow{2}{*}{\begin{tabular}[c]{@{}c@{}} \\ NN$_{1}$ \end{tabular}}                                                    & \multirow{2}{*}{\begin{tabular}[c]{@{}c@{}} \\ NN$_{1}$ \end{tabular}}                                                 \\ \cline{6-6}
                                                                  &                        &                                                                        &                                                                        &                                                                          & \begin{tabular}[c]{@{}c@{}}Likelihood \\ Estimation \cite{choi2019context} \end{tabular}  &                                                                        &                                                                     \\ \cline{3-8} 
                                                                         &                 & \begin{tabular}[c]{@{}c@{}}Uniform \\ Quantizer\end{tabular}           & \begin{tabular}[c]{@{}c@{}}GBI-based \\ Quantizer \cite{hanna2020distributed}\end{tabular}         & \begin{tabular}[c]{@{}c@{}}Uniform \\ Quantizer\end{tabular}             & \begin{tabular}[c]{@{}c@{}}Uniform \\ Quantizer\end{tabular}      & \begin{tabular}[c]{@{}c@{}}Uniform \\ Quantizer\end{tabular}           & \begin{tabular}[c]{@{}c@{}}Uniform \\ Quantizer\end{tabular}        \\ \hline
\multirow{2}{*}{\textbf{\begin{tabular}[c]{@{}c@{}}Server-based \\ Network\end{tabular}}} & NN$_{1}$ & \begin{tabular}[c]{@{}c@{}}Feature Vector\\ Concatenation\end{tabular} & \begin{tabular}[c]{@{}c@{}}Feature Vector\\ Concatenation\end{tabular} & \begin{tabular}[c]{@{}c@{}}Significant-Aware \\ Feature Selection \cite{shi2020communication} \end{tabular} & \begin{tabular}[c]{@{}c@{}}Weighted \\ Pooling Layer\end{tabular} & \begin{tabular}[c]{@{}c@{}}Feature Vector\\ Concatenation\end{tabular} & \begin{tabular}[c]{@{}c@{}}Selective \\ Retransmission\end{tabular} \\ \cline{2-8} 
                                                                                  & NN$_2$        & NN$_2$                                                                    & NN$_2$                                                                    & NN$_2$                                                                      & NN$_2$                                                               & NN$_2$                                                                    & NN$_2$                                                                 \\ \hline
\end{tabular}
}
\end{center}
\label{Table:network_structures}
\end{table*}

\begin{table*}[t]
\caption{The accuracy of the multi-view image classification task under different bit constraints. }
\begin{center}
\begin{threeparttable}
\setlength{\tabcolsep}{3mm}{
\begin{tabular}{c|ccc|ccc|cc}
\hline
%\toprule[1pt]
  & &{MNIST} &    & & {CIFAR} & &  \multicolumn{2}{c}{Tiny ImageNet (Top-1/Top-5)}     \\ %\cline{2-7} 
                      & 6 bits & 10 bits & 14 bits & 8 bits & 16 bits & 24 bits  & 256 bits & 512 bits \\ 
%\midrule[1pt]
\hline
NN-REG               & 95.93\%         & 97.49\%         &  97.78\%  & 85.39\%         & 89.74\%         &  90.14\%           & 49.17\%/74.65\%         &  50.29\%/76.03\%      \\
NN-GBI$^*$            & 96.62\%         & 97.79\%         & 98.02\%  & 86.37\%         & 89.69\%         & 90.30\%         & ---         &  ---      \\
eSAFS                 & 96.97\%         & 97.87\%         & 98.05\%  & 84.50\%         & 88.22\%         & 90.07\%           & 47.60\%/73.25\%         &  49.63\%/74.88\%  \\
CAFS            & 94.14\%         & 97.43\%         & 97.42\%        & 82.83\%         & 89.75\%         & 90.34\%           & 46.24\%/71.69\%         &  47.83\%/73.12\%     \\ 
%\midrule[1pt]
\hline
VDDIB (ours)          & 97.08\%         & 97.82\%        & 98.06\%  & 85.95\%         & 90.17\%        & 90.78\%            & 48.89\%/74.59\%         &  50.37\%/75.64\%   \\
VDDIB-SR (T=2) (ours) & \textbf{97.13\%}         & \textbf{98.13\%}         & \textbf{98.22\%}  & \textbf{87.61\%}         & \textbf{90.53\%}         & \textbf{90.94\%}           & \textbf{49.96\%}/\textbf{74.88\%}       & \textbf{51.03\%}/\textbf{76.06\%}      \\
%\bottomrule[1pt]
\hline
\end{tabular}}
\begin{tablenotes}
\item[$*$]{\scriptsize The GBI quantization algorithm is computationally prohibitive when the number of bits is too large.}
\end{tablenotes}
\end{threeparttable}

\end{center}
\label{Table:image_classification}
\end{table*}

\begin{table}[t]
%\caption{Global caption}
\caption{The accuracy of the multi-view object recognition task under different bit constraints.}
\centering
\resizebox{0.495\textwidth}{!}{
\begin{tabular}{c|ccc|ccc}
\hline
 & \multicolumn{3}{c|}{WRGBD} & \multicolumn{3}{c}{ModelNet40}           \\
                      & 12 bits & 15 bits & 18 bits & 120 bits & 240 bits & 360 bits \\ 
\hline
NN-REG                &93.97\%          &96.74\%       &98.50\%  &87.50\%          &88.25\%       &89.03\%       \\
NN-GBI\tnote{$*$}     &95.92\%          &98.62\%           & 98.82\%    &88.82\%         &---           & ---      \\
eSAFS                 &90.45\%         & 95.57\%        & 99.12\%  &85.88\%          & 87.87\%        & 89.50\%        \\
CAFS                  &86.65\%          &94.51\%          & 97.33\%  &86.75\%          &89.56\%          & 90.67\%        \\ 
%\midrule[1pt]
\hline
VDDIB (ours)          &94.93\%          & 97.47\%        & 98.63\%  &89.25\%          & 90.03\%        & 90.75\%        \\
VDDIB-SR (T=2) &\textbf{96.78\%}          & \textbf{99.30\%}         & \textbf{99.45\%}  &\textbf{90.25\%}          & \textbf{91.31\%}         & \textbf{91.62\%}       \\ 
\hline
\end{tabular}
}
\label{Table:shape_recognition}
\end{table}

\begin{table}[t]
\caption{The accuracy and cost of the server-based inference method.}
\centering
\resizebox{0.49\textwidth}{!}{
\begin{tabular}{c|c|c|c|c|c}
\hline
         & MNIST   & CIFAR-10 & Tiny-ImageNet                                                              & WRGBD   & ModelNet40 \\ \hline
Accuracy & 98.6\% & 91.5\%  & \begin{tabular}[c]{@{}c@{}}51.5\% (Top-1) \\ 76.9\% (Top-5)\end{tabular} & 99.8\% & 92.0\%    \\ \hline
Comm. cost & 0.34 KB & 2.31 KB  & 9.14 KB & 4.05 KB & 133 KB \\ \hline
\end{tabular}
}
\label{Table:server-based_inference}
\end{table}

\begin{table}[t]
\caption{The average accuracy of the single-view inference on different tasks.}
\centering
\resizebox{0.49\textwidth}{!}{
\begin{tabular}{c|c|c|c|c|c}
\hline
         & MNIST   & CIFAR-10 & Tiny-ImageNet                                                              & WRGBD   & ModelNet40 \\ \hline
Accuracy & 92.69\% & 83.16\%  & \begin{tabular}[c]{@{}c@{}}40.84\% (Top-1) \\ 65.89\% (Top-5)\end{tabular} & 99.16\% & 88.33\%    \\ \hline
\end{tabular}
}
\label{single_view_inference}
\end{table}

\subsubsection{Neural Network Architecture}
In the edge inference system, each raw input sample is passed through the first part of the network (denoted as $\text{NN}_{1}$) to extract the view-specific features, and the multiple features are collected for the processing of the remaining part of the network (denoted as $\text{NN}_{2}$) to output the inference result.
Generally, we select a convolutional (Conv) neural network as the backbone of $\text{NN}_{1}$, and $\text{NN}_{2}$ is constructed by several fully-connected (FC) layers.
More details about the structures of $\text{NN}_{1}$ and $\text{NN}_{2}$ are summarized in Table \ref{Table:NN1_NN2_structure}.

The architectures of different methods are presented in Table \ref{Table:network_structures}, where most of the methods adopt a uniform quantizer to discretize the output of NN$_1$ for efficient transmission.
In particular, the NN-GBI method adopts the GBI algorithm \cite{hanna2020distributed} to design a non-uniform quantizer, and the server-based inference scheme transmits the raw input views to the edge server via data-oriented communication.
After receiving the extracted features, the NN-REG, NN-GBI, and VDDIB methods use vector concatenation for feature aggregation.
The CAFS method leverages a weighted pooling layer to fuse the received features, where the weights are determined by the likelihood estimation module \cite{choi2019context}.
Besides, the eSAFS method performs signification-aware feature selection \cite{shi2020communication} to identify the most informative feature maps for transmission.
Our proposed VDDIB-SR leverages a selective retransmission mechanism to collect the features, where the attention module adopts two fully-connected layers to output the attention scores.

\subsubsection{Metrics}
We investigate the rate-relevance tradeoff of different multi-device edge inference schemes, where the rate is measured by the total number of bits ($R_{\text{sum}}$) to be transmitted from the edge devices to the edge server, and the relevance is measured by the classification accuracy.
In particular, we neglect the feedback signaling cost in the VDDIB-SR and eSAFS \cite{singhal2020communication} methods due to the much lower communication overhead compared with that of the forward transmission, as well as the abundant radio resources for the downlink transmission.
In the training process, we use the cross-entropy loss to measure the difference between the predicted results and the ground truths.
Besides, the proposed VDDIB-SR method selects the maximum predicted probability in (\ref{equ:confidence_score}) as the confidence score.

\subsection{Multi-View Image Classification}

This task aims to classify multiple views of an image to its target label.
We compare the inference performance given different constraints on the average received bits $R_{\text{sum}}$ over the test set.
Table \ref{Table:image_classification} shows the classification accuracy under different bit constraints, where the proposed VDDIB method achieves a comparable or better accuracy compared with the baselines.
This implies the effectiveness of the IB and DIB frameworks for task-oriented communication in multi-device edge inference systems.
Besides, the accuracy achieved by the VDDIB-SR method is higher than that of the VDDIB method.
This is attributed to the selective retransmission mechanism that eliminates the redundancy among the encoded features.
For better comparison, the average accuracy of single-view inference is reported in Table \ref{single_view_inference}. It illustrates that every single view is not sufficient to carry out the inference task with a good accuracy, which demonstrates the advantages of cooperative inference.
Finally, compared with the server-based inference scheme shown in Table \ref{Table:server-based_inference}, our proposed methods can greatly reduce the communication overhead with little drop in performance.

\begin{figure*}[t]
\centering

\subfloat[MNIST dataset.]{
\centering
\includegraphics[width=0.35\textwidth]{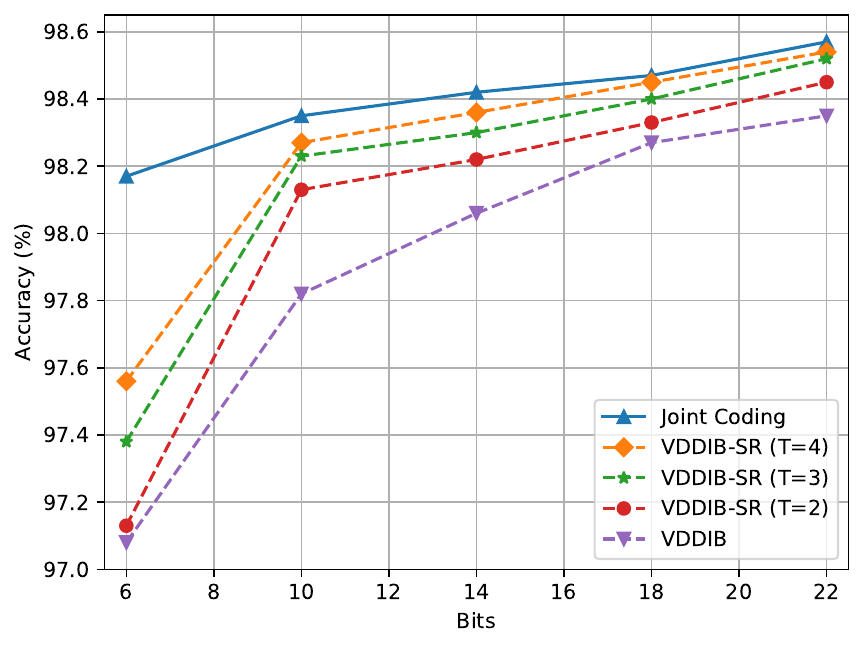}
}
\subfloat[WRGBD dataset.]{
\centering
\includegraphics[width=0.35\textwidth]{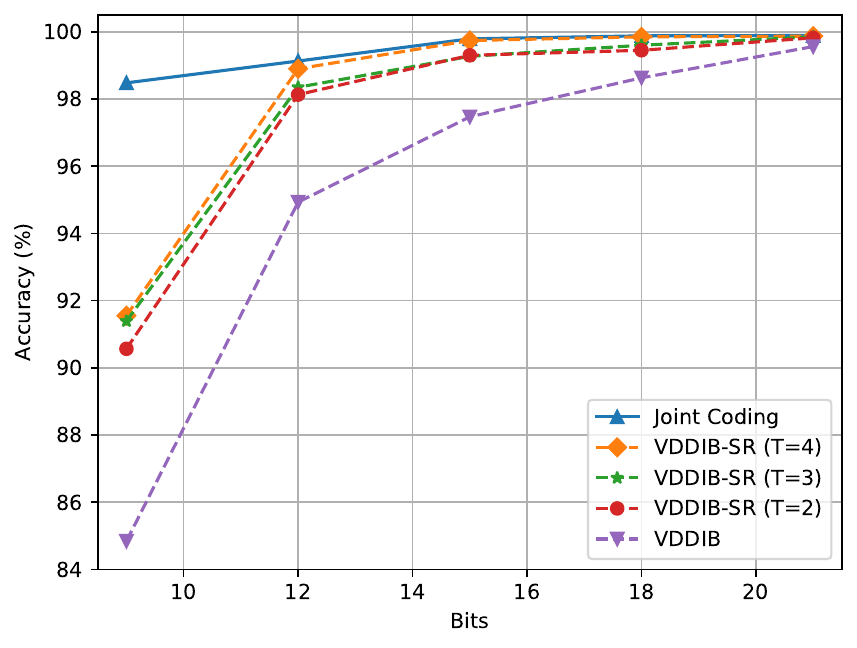}
}

\caption{Impact of the maximum transmission attempts $T$ on the rate-relevance tradeoff.}
\label{Fig:Impact_of_T}

\end{figure*}

\begin{figure*}[t]
\centering
\begin{minipage}[t]{0.35\textwidth}
\centering
\includegraphics[width=1\linewidth]{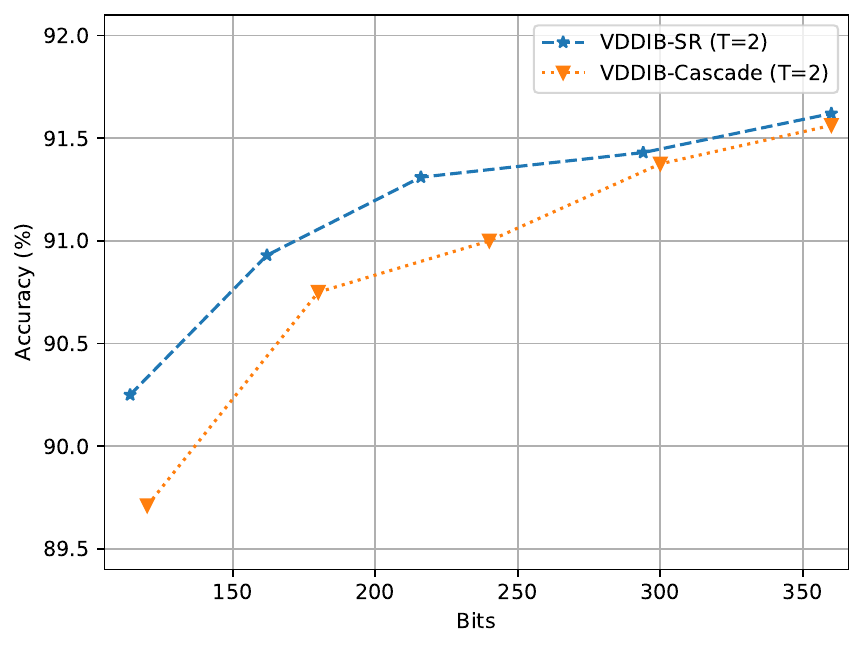}
%\caption{Impact of the attention module on the rate-relevance tradeoff.}
\caption{Accuracy as a function of communication cost on the ModelNet40 classification task.}
\label{Fig:ablation_attention_module}
\end{minipage}
\hspace{.1in}
\begin{minipage}[t]{0.35\textwidth}
\centering
\includegraphics[width=1\linewidth]{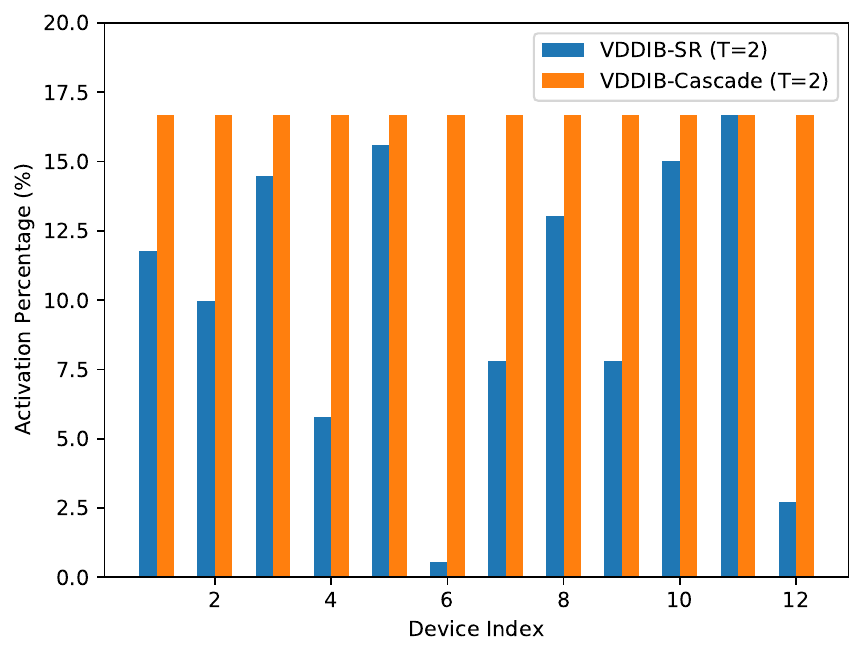}
\caption{The activation percentages of the edge devices in the retransmission.}
\label{Fig:activation percentages}
\end{minipage}
\end{figure*}

\subsection{Multi-View Object Recognition}
\label{Sec_shape_reco}
This task aims to recognize the 3D object based on a collection of perceived 2D views.
Consider that there are twelve devices with distinct views in the multi-device cooperative edge inference system. 
We compare the inference performance given different rate constraints.
Table \ref{Table:shape_recognition} summarizes the inference accuracy for the multi-view object recognition task. 
We see that our methods consistently outperform the baselines under all rate constraints.
This is because the proposed communication schemes are capable of discarding the task-irrelevant information.
Besides, with the selective retransmission mechanism, the VDDIB-SR method secures further performance enhancement than the VDDIB method. 
Besides, compared with the single-view inference results shown in Table \ref{single_view_inference}, the proposed VDDIB-SR method can effectively leverage the correlation among multiple views for better accuracy.
The performance and communication cost of the server-based inference method is reported in Table \ref{Table:server-based_inference}. 
Although this method achieves a slight performance gain, it induces overwhelming communication overhead, leading to intolerable latency.
%compared with the cooperative edge inference methods
Specifically, if the multi-device cooperative inference system is supported by the 3GPP narrowband Internet of Things (NB-IoT) standard with an 18 KB/s uplink rate \cite{zayas20173gpp_iot}, the data-oriented communication scheme has a communication latency of 560 ms on the ModelNet40 dataset, while the proposed VDDIB-SR method remarkably reduces the latency to 2.5 ms and maintains comparable classification accuracy.

% The server-based inference method that offloads raw input views to the edge server by data-oriented communication achieves an accuracy of around 92\%. 
% However, it induces around 120 KB communication overhead, leading to intolerable latency.

\subsection{Ablation Study}
{\color{black}
In this subsection, we evaluate the performance of (1) the maximum transmission attempts $T$ and (2) the attention module on the multi-device edge inference systems.
An ablation study that validates the effectiveness of the feature extraction step is deferred to Appendix \ref{Appendix:ablation}.}
\subsubsection{Impact of the maximum transmission attempts $T$}
We investigate the performance of the VDDIB-SR method with different values of $T$ on the multi-view image classification task.
In particular, we adopt a joint coding scheme as a baseline method for comparison, which assumes that one edge device can access all the extracted view-specific features and perform joint feature encoding before transmission.
The relationship between the achieved classification accuracy and the number of transmitted bits is shown in Fig. \ref{Fig:Impact_of_T}.
It can be observed that the performance of the VDDIB-SR method improves with $T$, while the joint source coding scheme achieves the best performance and upper bounds the VDDIB-SR method.
These results are consistent with the analysis in Remark \ref{remark:VDDIB-SR1} that increasing the maximum transmission attempts $T$ could achieve a better rate-relevance tradeoff\footnote{Note that when $T$ is very large, the feedback signaling cost is non-negligible. Selecting an optimal $T$ that reduces the overall communication costs while maintaining the inference performance requires a dedicated communication protocol to manage the data flow, which is an interesting future research but beyond the scope of this paper.}, and the joint coding scheme outperforms the distributed coding methods.

% {\color{blue} 
% It can be observed that the joint {source} coding scheme achieves the best result, and the performance of the VDDIB-SR method improves with $T$.
% These results are consistent with the analysis in Remark \ref{remark:VDDIB-SR1} that increasing the maximum transmission attempts $T$ could achieve a better rate-relevance tradeoff\footnote{Note that when $T$ is sufficiently large, the feedback signaling cost is not neglectable. Selecting an optimal $T$ that reduces the overall communication costs while maintaining the inference performance requires a dedicated communication protocol to manage the data flow, which is an interesting problem for future research. However, this problem is out of the scope of this paper.}.
% }

%It can be observed that the performance of the VDDIB-SR method improves with $T$, while the joint {source} coding scheme achieves the best performance and upper bound the VDDIB-SR method.
%These results are consistent with the analysis in Remark \ref{remark:VDDIB-SR1} that increasing the maximum transmission attempts $T$ could achieve a better rate-relevance tradeoff\footnote{Note that when $T$ is sufficiently large, the feedback signaling cost is not neglectable. Selecting an optimal $T$ that reduces the overall communication costs while maintaining the inference performance requires a dedicated communication protocol to manage the data flow, which is an interesting problem for future research. However, this problem is out of the scope of this paper.}, and the joint coding scheme outperforms the distributed coding methods.

\subsubsection{Impact of the attention module}
%To evaluate the performance of the attention module on communication overhead reduction, we select a baseline method denoted as VDDIB-Cascade for comparison, {\color{red}which activates all the devices in retransmission without the attention module.}
To evaluate the performance of the attention module on communication overhead reduction, we select a baseline method denoted as VDDIB-Cascade for comparison, {\color{black}which activates all the devices to transmit their features in each retransmission round before terminating the retransmission process.}
%{\color{red}which removes the attention module from the VDDIB-SR method and activates all the devices in retransmission.}
We investigate the performance of these two methods on the ModelNet40 dataset.
Specifically, the transmission attempt $T$ is set to 2.
As shown in Fig. \ref{Fig:ablation_attention_module}, the VDDIB-SR method achieves a better rate-relevance tradeoff compared with the VDDIB-Cascade method, which demonstrates the effectiveness of the attention module.
In particular, Fig. \ref{Fig:activation percentages} depicts the activation percentages of the edge devices in the retransmission, where the VDDIB-SR and VDDIB-Cascade methods induce 216 bits and 240 bits overhead, respectively.
Both methods achieve an accuracy of around 91.1\%, and the communication cost reduction is attributed to the attention module that can schedule the edge devices with the most informative features to retransmit.

\section{Conclusions}
\label{Sec:Conclusion}

This work investigated task-oriented communication for multi-device cooperative edge inference, where a group of low-end edge devices transmit the task-relevant features to an edge server for aggregation and processing. 
Our proposed methods are built upon the IB principle and the DIB framework for feature extraction and distributed feature encoding, respectively.
Compared with traditional data-oriented communication, our methods substantially reduce the communication overhead, thus enabling low-latency cooperative inference. 
Compared with existing methods for feature encoding in cooperative inference, we provide a theoretical framework to characterize the rate-relevance tradeoff, supported by pragmatic approaches for local feature extraction and distributed feature encoding, which consistently outperform baseline methods.

This study demonstrated the effectiveness of task-oriented communication for multi-device cooperative edge inference, which serves for the downstream task rather than for data reconstruction.
The shifted objective of the communication system calls for new design tools.
The IB-based frameworks, variational approximations, and retransmission mechanisms are promising candidates.
Given the above, it is interesting to further investigate task-oriented communication strategies for other edge AI systems and applications, e.g., to design tighter bounds in variational approximations, to develop effective coding schemes for dynamic conditions, and to apply the IB-based methods to decentralized networks \cite{reviewer_Moldoveanu2}, etc.

\appendices

\section{Proof of Remark \ref{remark:minimal_sufficient_feature}}

\label{Appendix:prove_minimal_sufficient_feature}

We prove that the extracted features $(Z_{1},\ldots,Z_{K})$ that satisfy the minimality and sufficiency conditions in (\ref{equ:MNI}) are conditionally independent given $Y$.
% \begin{lemma}
% \label{lemma:y_q_markov}
% For any joint distribution $p(\bm{x}_{1:K},\bm{y})$, there exists a random variable $Q$ that is independent of $Y$ and satisfies $p(\bm{x}_{1:K},\bm{y},\bm{q}) = p(\bm{y})p({\bm{q}})\prod_{k=1}^{K}p(\bm{x}_{k}|\bm{y},\bm{q})$.
% \end{lemma}
% \begin{proof}
% A trivial solution that satisfies $p(\bm{x}_{1:K},\bm{y},\bm{q}) = p(\bm{y})p({\bm{q}})\prod_{k=1}^{K}p(\bm{x}_{k}|\bm{y},\bm{q})$ is that $p({\bm{q}})= p(\bm{x}_{1:K}|\bm{y})$.
% \end{proof}
Define a random variable $Q$ with distribution $p(\bm{q})$ that satisfies the following equality\footnote{The random variable $Q$ exists for any task by trivially defining $Q := X_{1:K}$.}:
\begin{equation}
\label{equ:definition_Q}
    p(\bm{x}_{1:K},\bm{y},\bm{q}) = p(\bm{y}|\bm{q})p({\bm{q}})\prod_{k=1}^{K}p(\bm{x}_{k}|\bm{y},\bm{q}).
\end{equation}
The random variables $Y$, $Q$, $X_{k}$, $Z_{k}$ constitute the following Markov chain since the extracted feature $Z_{k}$ is independent of $(Y,Q)$ given the local data $X_{k}$:
\begin{equation}
\label{equ:y_q_x_z}
    (Y,Q) \leftrightarrow X_{k} \leftrightarrow Z_{k}. \ k \in \{1,\ldots,K\}.
\end{equation}
With the data processing inequality, we have $I(Y,Q;Z_{k}) \leq I(X_{k},Z_{k})$.
Suppose the extracted features satisfy the minimality and sufficiency in (\ref{equ:MNI}), which implies $I(X_{k};Z_{k}) = I(Y;Z_{k})$, $k \in \{1,\ldots,K\}$.
By incorporating the chain rule of mutual information, i.e., $I(Y;Z_{k}) \leq I(Y,Q;Z_{k})$, we get $I(Y,Q;Z_{k}) = I(Y;Z_{k})$, which shows that the random variables $Q,Y,Z_{k}$ constitute the Markov chain as follows:
\begin{equation}
\label{equ:markov_q_y_z}
     Q \leftrightarrow Y \leftrightarrow Z_{k}, \ k \in \{1,\ldots,K\}.
\end{equation}
As each extracted feature $\bm{z}_{k}$ only depends on the corresponding observation $\bm{x}_{k}$, the conditional distribution of $(X_{1:K},Z_{1:K})$ given $(Y,Q)$ is as follows:
\begin{align}
p(\bm{x}_{1:K},\bm{z}_{1:K}|\bm{y},\bm{q}) = & \prod_{k=1}^{K} p(\bm{x}_{k}|\bm{y},\bm{q})p(\bm{z}_{k}|\bm{x}_{k}),   \label{align:prop2_def1}  \\
= & \prod_{k=1}^{K} p(\bm{x}_{k},\bm{z}_{k}|\bm{y},\bm{q}). \label{align:prop2_def2}
\end{align}
where (\ref{align:prop2_def1}) is due to (\ref{equ:definition_Q}), and (\ref{align:prop2_def2}) is derived from the Markov chains in (\ref{equ:y_q_x_z}).
By integrating out $\bm{x}_{1:K}$ in (\ref{align:prop2_def2}), the conditional distribution of $Z_{1:K}$ given $(Y,Q)$ is as follows:
\begin{align}
p(\bm{z}_{1:K}|\bm{y},\bm{q}) = & \prod_{k=1}^{K} p(\bm{z}_{k}|\bm{y},\bm{q}),   \label{align:prop2_def0}  \\
= & \prod_{k=1}^{K} p(\bm{z}_{k}|\bm{y}), \label{align:prop2_def3}
\end{align}
where (\ref{align:prop2_def3}) is due to the Markov chains in (\ref{equ:markov_q_y_z}).
By integrating out $\bm{q}$ in the left-hand side of (\ref{align:prop2_def3}), we have $p(\bm{z}_{1:K}|\bm{y}) =  \prod_{k=1}^{K} p(\bm{z}_{k}|\bm{y})$, which means that the features $Z_{1:K}$ are conditionally independent given the target variable $Y$.

\begin{figure*}
\normalsize
\begin{equation}
\label{equ_appendix_vib}
\begin{aligned}
 \mathcal{L}_{\mathrm{IB},k}(\gamma) = & \mathbf{E}_{p(\bm{x}_{k},\bm{y})}\left\{ \mathbf{E}_{p_{\bm{\theta}_{k}}({\bm{z}_{k}}|\bm{x}_{k})} [ - \log p(\bm{y}|\bm{z}_{k}) ]  + \gamma D_{\mathrm{KL}}(p_{\bm{\theta}_{k}}(\bm{z}_{k}|\bm{x}_{k})\|p(\bm{z}_{k})) \right\}, \\
 = & \underbrace{ \mathbf{E}_{p(\bm{x}_{k},\bm{y})}\left\{ \mathbf{E}_{p_{\bm{\theta}_{k}}({\bm{z}_{k}}|\bm{x}_{k})} [ - \log p_{\bm{\varphi}_{k}}(\bm{y}|\bm{z}_{k}) ]  + \gamma D_{\mathrm{KL}}(p_{\bm{\theta}_{k}}(\bm{z}_{k}|\bm{x}_{k})\|r_{k}(\bm{z}_{k})) \right\} }_{\mathcal{L}_{\mathrm{VIB},k}(\gamma;\bm{\theta}_{k},\bm{\varphi}_{k})} \\
 & - \mathbf{E}_{p(\bm{z}_{k})} \left\{ D_{\mathrm{KL}}(p(\bm{y}|\bm{z}_{k})\|p_{\bm{\varphi}_{k}}(\bm{y}|\bm{z}_{k}))\right\} - \gamma  D_{\mathrm{KL}}(p(\bm{z}_{k})\|r_{k}(\bm{z}_{k})) .
 %& - \mathbf{E}_{p(\bm{x}_{k},\bm{z}_{k})} \left\{ D_{\mathrm{KL}}(p(\bm{y}|\bm{z}_{k})\|p_{\bm{\varphi}_{k}}(\bm{y}|\bm{z}_{k})) + \gamma  D_{\mathrm{KL}}(p(\bm{z}_{k})\|r_{k}(\bm{z}_{k})) \right\}.
\end{aligned}
\end{equation}
\hrulefill
\end{figure*}

\begin{figure*}
\normalsize
\begin{equation}
\label{equ_appendix_vddib}
\begin{aligned}
    \mathcal{L}_{\mathrm{DDIB}}(\beta) = & \mathbf{E}\left\{ - \log p(\bm{y}|\bm{u}_{1:K}) + \beta \sum_{k=1}^{K} \left[  - \log p(\bm{y}|\bm{u}_{k}) + R_{\mathrm{bit}}(\bm{u_{k}}) \right]   \right\}, \\
    = & \underbrace{ \mathbf{E}\left\{ - \log p_{\bm{\psi_{0}}}(\bm{y}|\bm{u}_{1:K}) + \beta \sum_{k=1}^{K} \left[  - \log p_{\bm{\psi}_{k}}(\bm{y}|\bm{u}_{k}) + R_{\mathrm{bit}}(\bm{u_{k}}) \right]  \right\}}_{\mathcal{L}_{\mathrm{VDDIB}}(\beta;\bm{\phi},\bm{\psi})} \\
    & - \mathbf{E}\left\{ D_{\mathrm{KL}}(p(\bm{y}|\bm{u}_{1:K})\|p_{\bm{\psi_{0}}}(\bm{y}|\bm{u}_{1:K})) +  \beta \sum_{k=1}^{K} D_{\mathrm{KL}}(p(\bm{y}\|\bm{u}_{k})|p_{\bm{\psi}_{k}}(\bm{y}|\bm{u}_{k})) \right\}. \\
\end{aligned}
\end{equation}
\hrulefill
\end{figure*}

\begin{figure*}
\normalsize
\begin{align}
    \min_{\bm{P}} \mathcal{L}_{\mathrm{DDIB}}(\beta)  \leq &  \min_{ \bm{P}}  \ \mathbf{E}\Bigg\{ \sum_{\tau=1}^{T} -\log p(\bm{y}|\{\bm{u}_{1:K,t}\}_{t=1}^{\tau})   +  \left. \beta \sum_{k=1}^{K}\left[ -\log p(\bm{y}|\bm{u}_{k}) +\sum_{t=1}^{T} R_{\text{bit}}(\bm{u}_{k,t})\right]  \right\} , \label{align:DDIB_with_variational_approximation}\\
    \leq & \min_{ \bm{P_{T}}} \ \mathcal{L}_{\mathrm{VDDIB-SR}}\left(\beta,T;\bm{\widetilde{\phi}},\widetilde{\bm{\psi}},\{\bm{\psi}_{k}\}_{k=1}^{K}\right), \label{align:DDIB_parameter_space}
\end{align}
\hrulefill
\end{figure*}

\section{Derivation of the Variational Upper Bound}

\subsection{Variational Information Bottleneck}
\label{Appendix:vib}

Recall that the IB objective in (\ref{equ:vib}) has the form $\mathcal{L}_{\mathrm{IB},k}(\gamma) = H(Y|Z_{k})+\gamma I(X_{k};Z_{k})$ at device $k \in \{1,\ldots,K\}$. Writing it out in full with the conditional distribution $p_{\bm{\theta}_{k}}(\bm{z}_{k}|\bm{x})$ and the variational distributions $p_{\bm{\varphi}_{k}}(\bm{y}|\bm{z}_{k})$, $r_{k}(\bm{z}_{k})$, the derivation is shown in (\ref{equ_appendix_vib}).
% it becomes:
% \begin{equation*}
% \begin{aligned}
%  \mathcal{L}_{\mathrm{IB},k}(\gamma) = & \mathbf{E}_{p(\bm{x}_{k},\bm{y})}\left\{ \mathbf{E}_{p_{\bm{\theta}_{k}}({\bm{z}_{k}}|\bm{x}_{k})} [ - \log p(\bm{y}|\bm{z}_{k}) ]  + \gamma D_{\mathrm{KL}}(p(\bm{z}_{k}|\bm{x}_{k})\|p(\bm{z}_{k})) \right\}, \\
%  = & \underbrace{ \mathbf{E}_{p(\bm{x}_{k},\bm{y})}\left\{ \mathbf{E}_{p_{\bm{\theta}_{k}}({\bm{z}_{k}}|\bm{x}_{k})} [ - \log p_{\bm{\varphi}_{k}}(\bm{y}|\bm{z}_{k}) ]  + \gamma D_{\mathrm{KL}}(p(\bm{z}_{k}|\bm{x}_{k})\|p(\bm{z}_{k})) \right\} }_{\mathcal{L}_{\mathrm{VIB},k}(\gamma;\bm{\theta}_{k},\bm{\varphi}_{k})} \\
%  & - \mathbf{E}_{p(\bm{x}_{k},\bm{y})} \left\{ D_{\mathrm{KL}}(p(\bm{y}|\bm{z}_{k})\|p_{\bm{\varphi}_{k}}(\bm{y}|\bm{z}_{k})) + \gamma  D_{\mathrm{KL}}(p(\bm{z}_{k})\|r_{k}(\bm{z}_{k})) \right\}.
% \end{aligned}
% \end{equation*}
$\mathcal{L}_{\mathrm{VIB},k}(\gamma;\bm{\theta_{k},\varphi_{k}})$ in the  formulation is the VIB objective function in (\ref{equ:vib}).
As the KL-divergence is non-negative, the VIB objective function is an upper bound of the IB objective function.

\subsection{Variational Distributed Deterministic Information Bottleneck}
\label{Appendix:VDDIB}

Revisit the DDIB objective function $\mathcal{L}_{\mathrm{DDIB}}(\beta) = H\left(Y | U_{1: K}\right)+\beta \sum_{k=1}^{K}\left[H\left(Y | U_{k}\right)+R_{\mathrm{bit}}\left(U_{k}\right)\right]$ in (\ref{equ:ddib}). 
Writing it out with the variational distributions $p_{\bm{\psi_{0}}}(\bm{y}|\bm{u}_{1:K})$ and $p_{\bm{\psi}_{k}}(\bm{y}|\bm{u}_{k})$,
the derivation is shown in (\ref{equ_appendix_vddib}).
%learned distribution $p_{\bm{\theta}}(\bm{z}_{1:K},\bm{y})$  and the encoding functions $\bm{u}_{k}=\bm{\mathcal{Q}}\left(\bm{f_{\phi_{k}}}\left(\bm{z}_{k}\right)\right)$, $k \in \{1,\ldots,K\}$, the derivation is shown in (\ref{equ_appendix_vddib}).
%shown in
% we have:
% \begin{equation*}
% \begin{aligned}
%     \mathcal{L}_{\mathrm{DDIB}}(\beta) = & \mathbf{E}_{p_{\bm{\theta}}(\bm{z}_{1:K},\bm{y})}\left\{ - \log p(\bm{y}|\bm{u}_{1:K}) + \beta \sum_{k=1}^{K} \left[  - \log p(\bm{y}|\bm{u}_{k}) + R_{\mathrm{bit}}(\bm{u_{k}}) \right]   \right\}, \\
%     = & \underbrace{ \mathbf{E}_{p_{\bm{\theta}}(\bm{z}_{1:K},\bm{y})}\left\{ - \log p_{\bm{\psi_{0}}}(\bm{y}|\bm{u}_{1:K}) + \beta \sum_{k=1}^{K} \left[  - \log p_{\bm{\psi}_{k}}(\bm{y}|\bm{u}_{k}) + R_{\mathrm{bit}}(\bm{u_{k}}) \right]  \right\}}_{\mathcal{L}_{\mathrm{VDDIB}}(\beta;\bm{\phi},\bm{\psi})} \\
%     & - \mathbf{E}_{p_{\bm{\theta}}(\bm{z}_{1:K},\bm{y})}\left\{ D_{\mathrm{KL}}(p(\bm{y}|\bm{u}_{1:K})\|p_{\bm{\psi_{0}}}(\bm{y}|\bm{u}_{1:K})) +  \beta \sum_{k=1}^{K} D_{\mathrm{KL}}(p(\bm{y}\|\bm{u}_{k})|p_{\bm{\psi}_{k}}(\bm{y}|\bm{u}_{k})) \right\}. \\
% \end{aligned}
% \end{equation*}
As the KL-divergence is non-negative, the VDDIB objective function is a variational upper bound of the DDIB objective function.

\section{Derivation of Remark \ref{remark:VDDIB-SR1}}
\label{appendix:vddib-sr}

In this part, we detail the derivation that the minimum of $\mathcal{L}_{\mathrm{DDIB}}(\beta)$ over $\bm{P} = \{p(\bm{u}_{1:K}|\bm{z}_{1:K})\}$ lower bounds the VDDIB-SR objective. 
%for any integer $T$.
%In this part, we aim to detail the derivation of the inequality in (\ref{equ:remark_VDDIB_SR_upper_bound}).
%Based on the definition in (\ref{equ:VDDIB-SR_DIB_gap}), it is equivalent to validate the following inequality to show (\ref{equ:remark_VDDIB_SR_upper_bound}):
%Then, by combining (\ref{equ:remark_VDDIB_SR_upper_bound}) and (\ref{equ:VDDIB-SR_DIB_gap}), we try to show that
%defined in (\ref{equ:PT_defined})
% \begin{equation}
% \label{equ:prove_looser_bound}
%     \min_{\bm{\widetilde{\phi}},\widetilde{\bm{\psi}},\{\bm{\psi}_{k}\}_{k=1}^{K}} \mathcal{L}_{\mathrm{VDDIB-SR}}\left(\beta,T;\bm{\widetilde{\phi}},\widetilde{\bm{\psi}},\{\bm{\psi}_{k}\}_{k=1}^{K}\right) \geq \min_{\bm{P}_{T}}  \mathcal{L}_{\mathrm{DIB}}(\beta).
% \end{equation}
Revisit the DDIB objective $\mathcal{L}_{\mathrm{DDIB}}(\beta) = H\left(Y | U_{1: K}\right)+\beta \sum_{k=1}^{K}\left[H\left(Y | U_{k}\right)+R_{\text{bit}}(U_{k}))\right]$ in (\ref{equ:dib}).
It satisfies the inequalities (\ref{align:DDIB_with_variational_approximation}) and (\ref{align:DDIB_parameter_space}),
%and we have:
% \begin{align}
%     & \min_{\bm{P}} \mathcal{L}_{\mathrm{DDIB}}(\beta) \\
%     & \leq  \min_{ \bm{P}}  \ \mathbf{E}_{p_{\bm{\theta}}(\bm{z}_{1:K},\bm{y})}\Bigg\{\mathbf{E}_{p(\bm{u}_{1:K}|\bm{z}_{1:K})} \Bigg\{ \sum_{\tau=1}^{T} -\log p_{\bm{\widetilde{\psi}}_{\tau}}(\bm{y}|\{\bm{u}_{1:K,t}\}_{t=1}^{\tau})  \notag \\
%     & \quad +  \left. \left. \beta \sum_{k=1}^{K}\left[ -\log p_{\bm{\psi}_{k}}(\bm{y}|\bm{u}_{k}) +\sum_{t=1}^{T} R_{\text{bit}}(\bm{u}_{k,t})\right]  \right\} \right\}, \label{align:DDIB_with_variational_approximation}\\
%     \leq & \min_{ \bm{P_{T}}} \ \mathcal{L}_{\mathrm{VDDIB-SR}}\left(\beta,T;\bm{\widetilde{\phi}},\widetilde{\bm{\psi}},\{\bm{\psi}_{k}\}_{k=1}^{K}\right), \label{align:DDIB_parameter_space}
% \end{align}
where (\ref{align:DDIB_with_variational_approximation}) is due to the variational approximations, and (\ref{align:DDIB_parameter_space}) holds since the family of encoding functions parameterized by $\bm{\widetilde{\phi}}$ is a subset of $\bm{P}$.

\begin{figure}
    \centering
    \includegraphics[width=1\linewidth]{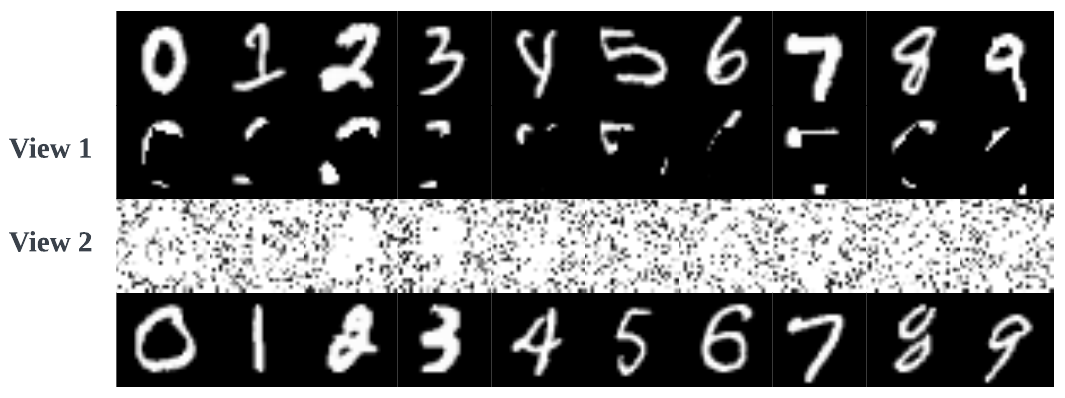}
    \caption{The corrupted two-view MNIST dataset}
    \label{fig:data_corrupted_two_view}
\end{figure}

% \usepackage{multirow}
\begin{table}[]
\caption{The neural network structure for corrupted two-view MNIST classification.}
\centering
\resizebox{0.5\textwidth}{!}{
\begin{tabular}{c|cc}
\hline
                                                                             & \multicolumn{1}{c|}{Our method}                            & D-VIB \cite{DVIB}          \\ \hline
\multirow{3}{*}{\begin{tabular}[c]{@{}c@{}}Feature\\ Extraction\end{tabular}} & \multicolumn{1}{c|}{{[}Conv+ReLU+MaxPooling{]} $\times$ 2} & \multirow{3}{*}{---}                   \\ \cline{2-2}
                                                                             & \multicolumn{1}{c|}{FC + ReLU + Dropout}                   &                                       \\ \cline{2-2}
                                                                             & \multicolumn{1}{c|}{FC + ReLU}                             &                                       \\ \hline
\multirow{3}{*}{Encoder}                                                     & \multicolumn{1}{c|}{\multirow{3}{*}{FC + ReLU}}            & {[}Conv+ReLU+MaxPooling{]} $\times$ 2 \\ \cline{3-3} 
                                                                             & \multicolumn{1}{c|}{}                                      & FC + ReLU + Dropout                   \\ \cline{3-3} 
                                                                             & \multicolumn{1}{c|}{}                                      & {[}FC + ReLU{]} $\times$ 2             \\ \hline
Decoder                                                                      & \multicolumn{2}{c}{FC + ReLU}                                                                      \\ \hline
\end{tabular}
}
\label{table:ablation_structure}
\end{table}

\begin{figure*}[t]
\centering
\begin{minipage}[t]{0.35\textwidth}
\centering
\includegraphics[width=1\linewidth]{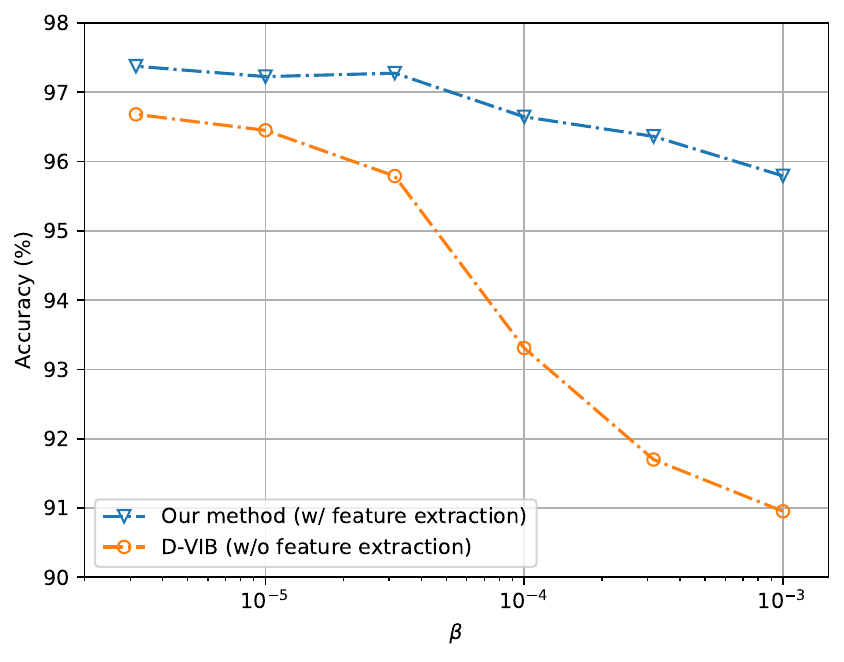}
%\caption{Impact of the attention module on the rate-relevance tradeoff.}
\caption{Accuracy as a function of $\beta$ in DIB for the corrupted two-view MNIST classification task.}
\label{Fig:appendix_accuracy}
\end{minipage}
\hspace{.1in}
\begin{minipage}[t]{0.355\textwidth}
\centering
\includegraphics[width=1\linewidth]{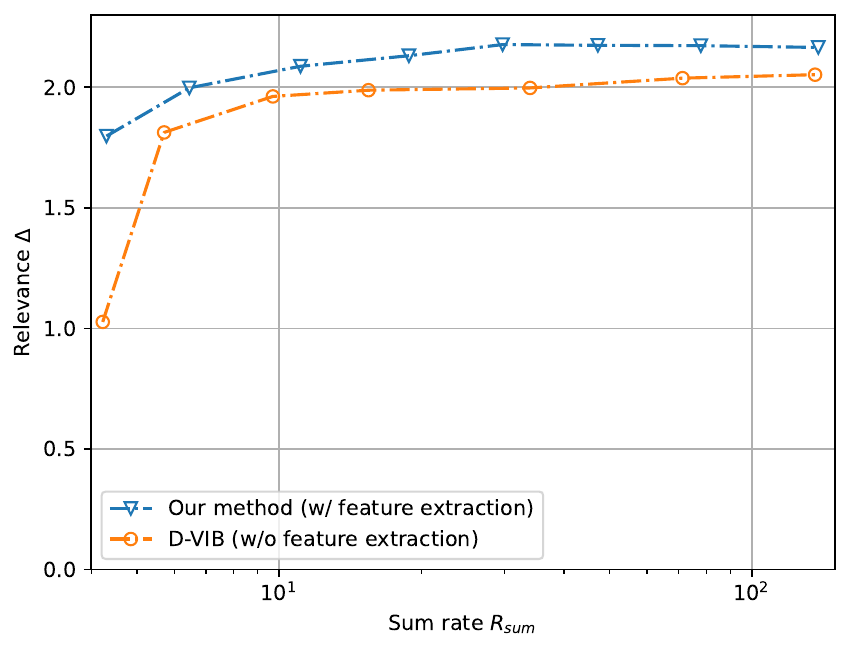}
\caption{Rate-relevance tradeoff for the corrupted two-view MNIST classification task.}
\label{Fig:appendix_tradeoff}
\end{minipage}
\end{figure*}

\section{Ablation Study on Feature Extraction}
\label{Appendix:ablation}
In this appendix, we evaluate the advantages of the added task-relevant feature extraction step on distributed coding.
The D-VIB method in \cite{DVIB} is selected as a baseline, which performs distributed coding on the raw input views without the feature extraction step.
%leverages variational distributed information bottleneck (D-VIB) loss to perform the  distributed coding on the raw input views without the feature extraction step.
%In particular, we select the D-VIB method in \cite{DVIB} as the baseline that performs the variational distributed coding on the raw input views without the feature extraction step.
For a fair comparison, we reproduce their results following the same experimental setup by Pytorch.
%follow the experimental setup in \cite{DVIB} for comparison.
%Since we reproduce their results on Pytorch
A corrupted two-view MNIST dataset for classification is generated for the two-view classification task, which is visualized in Fig. \ref{fig:data_corrupted_two_view}.
View 1 is generated by using a $15 \times 15$ mask to occlude part of the image.
View 2 is generated by adding independent random noise uniformly sampled from $[0, 3]$ to each pixel, and then the pixel values are truncated to $[0, 1]$.
We randomly select 50,000 two-view samples for model training and 20,000 samples for testing.
The neural network structures are presented in Table \ref{table:ablation_structure}, which consists of convolutional (Conv) layers, max-pooling layers, fully-connected (FC) layers, a dropout layer, and the rectified linear unit (ReLU) activation function.
Our proposed method utilizes the same neural network structure as D-VIB for a fair comparison.
Different from the baseline, we first optimize the layers for feature extraction based on Algorithm \ref{algorithm:VIB} and then train the encoder and decoder based on the variational approximation of the DIB loss function \cite{DVIB}.

% in (\ref{equ:dib})
%distributed information bottleneck (DIB) loss function in (\ref{}).
%the information bottleneck (IB) loss in (\ref{}) with parameter $\gamma = 10^{-5}$, and then train the encoder and decoder based on the distributed information bottleneck (DIB) loss function in (\ref{}).

Fig. \ref{Fig:appendix_accuracy} shows the classification accuracy given different values of $\beta$ in the DIB loss (\ref{equ:dib}).
Besides, we plot the achievable rate-relevance pairs in Fig. \ref{Fig:appendix_tradeoff} according to (\ref{equ:def_rate_relevance}) based on the neural network parameters.
Particularly, the conditional entropy of the target variable given the encoded features is estimated by the cross-entropy, and the marginal distributions of the encoded features are approximated by the variational distributions.
%The result is shown 
%and show the achievable rate-relevance pairs according to (\ref{equ:def_relevance}) and (\ref{equ:def_rate}) based on the neural network parameters.
%Particularly, the conditional entropy of the target variable given the encoded features is estimated by the cross-entropy, and the marginal distributions of the encoded features are approximated by the variational distributions.
We observe that our method achieves higher accuracy and a better rate-relevance tradeoff.
These can be explained by the advantage of the added feature extraction step, which discards the task-irrelevant information and allows the extracted features to be less correlated compared with the raw input views.

\ifCLASSOPTIONcaptionsoff
  \newpage
\fi



%\bibliographystyle{IEEEtran}

\bibliographystyle{IEEEtran}
%\bibliographystyle{ieeetr}
%\bibliography{/Users/yuanming/OneDrive/Paper/topics/Reference}
%\bibliography{ref}
\bibliography{IEEEabrv,ref}

\begin{thebibliography}{10}
\providecommand{\url}[1]{#1}
\csname url@samestyle\endcsname
\providecommand{\newblock}{\relax}
\providecommand{\bibinfo}[2]{#2}
\providecommand{\BIBentrySTDinterwordspacing}{\spaceskip=0pt\relax}
\providecommand{\BIBentryALTinterwordstretchfactor}{4}
\providecommand{\BIBentryALTinterwordspacing}{\spaceskip=\fontdimen2\font plus
\BIBentryALTinterwordstretchfactor\fontdimen3\font minus
  \fontdimen4\font\relax}
\providecommand{\BIBforeignlanguage}[2]{{%
\expandafter\ifx\csname l@#1\endcsname\relax
\typeout{** WARNING: IEEEtran.bst: No hyphenation pattern has been}%
\typeout{** loaded for the language `#1'. Using the pattern for}%
\typeout{** the default language instead.}%
\else
\language=\csname l@#1\endcsname
\fi
#2}}
\providecommand{\BIBdecl}{\relax}
\BIBdecl

\bibitem{tishby2000informationIB}
N.~Tishby, F.~C. Pereira, and W.~Bialek, ``The information bottleneck method,''
  in \emph{Proc. Annu. Allerton Conf. Commun. Control Comput.}, Monticello, IL,
  USA, Oct. 2000.

\bibitem{DVIB}
I.~E. Aguerri and A.~Zaidi, ``Distributed variational representation
  learning,'' \emph{IEEE Trans. Pattern Anal. Mach. Intell.}, vol.~43, no.~1,
  pp. 120--138, Jan. 2021.

\bibitem{computer_vision}
C.~Szegedy, V.~Vanhoucke, S.~Ioffe, J.~Shlens, and Z.~Wojna, ``Rethinking the
  inception architecture for computer vision,'' in \emph{Proc. IEEE Conf.
  Comput Vision Pattern Recognit.}, Las Vegas, NV, USA, Jun. 2016.

\bibitem{vr}
X.~Hou, S.~Dey, J.~Zhang, and M.~Budagavi, ``Predictive view generation to
  enable mobile 360-degree and {VR} experiences,'' in \emph{Proc. Morning
  Workshop VR AR Netw.}, Budapest, Hungary, Aug. 2018.

\bibitem{natureLP}
R.~Collobert and J.~Weston, ``A unified architecture for natural language
  processing: Deep neural networks with multitask learning,'' in \emph{Proc.
  Int. Conf. Mach. Learn.}, Helsinki, Finland, Jul. 2008.

\bibitem{shi2020communication}
Y.~Shi, K.~Yang, T.~Jiang, J.~Zhang, and K.~B. Letaief,
  ``Communication-efficient edge {AI}: Algorithms and systems,'' \emph{IEEE
  Commun. Surv. Tut.}, vol.~22, no.~4, pp. 2167--2191, Jul. 2020.

\bibitem{shao2020communication}
J.~Shao and J.~Zhang, ``Communication-computation trade-off in
  resource-constrained edge inference,'' \emph{IEEE Commun. Mag.}, vol.~58,
  no.~12, pp. 20--26, Dec. 2020.

\bibitem{reviewer_Moldoveanu1}
M.~Moldoveanu and A.~Zaidi, ``On in-network learning: a comparative study with
  federated and split learning,'' in \emph{Proc. Int. Workshop Signal Process.
  Adv. Wireless Commun. (SPAWC)}, Lucca, Italy, Sep. 2021.

\bibitem{self-driving_cars}
C.~Badue, R.~Guidolini, R.~V. Carneiro, P.~Azevedo, V.~B. Cardoso, A.~Forechi,
  L.~Jesus, R.~Berriel, T.~M. Paixao, F.~Mutz \emph{et~al.}, ``Self-driving
  cars: A survey,'' \emph{Expert Syst. Appl.}, p. 113816, Mar. 2020.

\bibitem{unlu2019deep_drone}
E.~Unlu, E.~Zenou, N.~Riviere, and P.-E. Dupouy, ``Deep learning-based
  strategies for the detection and tracking of drones using several cameras,''
  \emph{IPSJ Trans. Comput. Vision Appl.}, vol.~11, no.~1, pp. 1--13, Jul.
  2019.

\bibitem{zou2019collaborative_robotics}
D.~Zou, P.~Tan, and W.~Yu, ``Collaborative visual slam for multiple agents: A
  brief survey,'' \emph{Virtual Reality Intell. Hardware}, vol.~1, no.~5, pp.
  461--482, Oct. 2019.

\bibitem{zhou2019edge_edge_intell}
Z.~Zhou, X.~Chen, E.~Li, L.~Zeng, K.~Luo, and J.~Zhang, ``Edge intelligence:
  Paving the last mile of artificial intelligence with edge computing,''
  \emph{Proc. IEEE}, vol. 107, no.~8, pp. 1738--1762, Aug. 2019.

\bibitem{iccshao}
J.~Shao and J.~Zhang, ``Bottlenet++: An end-to-end approach for feature
  compression in device-edge co-inference systems,'' in \emph{Proc. IEEE Int.
  Conf. Commun. Workshop}, Dublin, Ireland, Jun. 2020.

\bibitem{shao2020branchy}
J.~Shao, H.~Zhang, Y.~Mao, and J.~Zhang, ``Branchy-gnn: A device-edge
  co-inference framework for efficient point cloud processing,'' in \emph{Proc.
  IEEE Int. Conf. Acoust. Speech Signal Process. (ICASSP)}, Toronto, Canada,
  Jun. 2021.

\bibitem{shao2021learning}
J.~Shao, Y.~Mao, and J.~Zhang, ``Learning task-oriented communication for edge
  inference: An information bottleneck approach,'' \emph{IEEE J. Sel. Area
  Commun.}, vol.~40, no.~1, pp. 197--211, Nov. 2021.

\bibitem{surveillance}
X.~Liu, W.~Liu, T.~Mei, and H.~Ma, ``A deep learning-based approach to
  progressive vehicle re-identification for urban surveillance,'' in
  \emph{Proc. Eur. Conf. Comput. Vision (ECCV)}, Amsterdam, Netherlands, Oct.
  2016.

\bibitem{strinati20216g_goal_oriented}
E.~C. Strinati and S.~Barbarossa, ``6{G} networks: Beyond shannon towards
  semantic and goal-oriented communications,'' \emph{Comput. Netw.}, vol. 190,
  p. 107930, May 2021.

\bibitem{zhu2020toward}
G.~Zhu, D.~Liu, Y.~Du, C.~You, J.~Zhang, and K.~Huang, ``Toward an intelligent
  edge: wireless communication meets machine learning,'' \emph{IEEE Commun.
  Mag.}, vol.~58, no.~1, pp. 19--25, Jan. 2020.

\bibitem{bottlenet}
A.~E. Eshratifar, A.~Esmaili, and M.~Pedram, ``Bottlenet: A deep learning
  architecture for intelligent mobile cloud computing services,'' in
  \emph{Proc. Int. Symp. Low Power Electron. Design (ISLPED)}, 2019, pp. 1--6.

\bibitem{dubois2021lossy}
Y.~Dubois, B.~Bloem-Reddy, K.~Ullrich, and C.~J. Maddison, ``Lossy compression
  for lossless prediction,'' in \emph{Proc. Int. Conf. Learn. Repr. Workshop on
  Neural Compression}, Vienna, Austria, May 2021.

\bibitem{jankowski2020wireless_Jankowski}
M.~{Jankowski}, D.~{Gündüz}, and K.~{Mikolajczyk}, ``Wireless image retrieval
  at the edge,'' \emph{IEEE J. Sel. Areas Commun.}, vol.~39, no.~1, pp.
  89--100, May 2021.

\bibitem{slepian1973noiseless_SW_Theorem}
D.~Slepian and J.~Wolf, ``Noiseless coding of correlated information sources,''
  \emph{IEEE Trans. Inf. Theory}, vol.~19, no.~4, pp. 471--480, Jul 1973.

\bibitem{hanna2020distributed}
O.~A. Hanna, Y.~H. Ezzeldin, T.~Sadjadpour, C.~Fragouli, and S.~Diggavi, ``On
  distributed quantization for classification,'' \emph{IEEE J. Sel. Areas Inf.
  Theory}, vol.~1, no.~1, pp. 237--249, Apr. 2020.

\bibitem{singhal2020communication}
M.~Singhal, V.~Raghunathan, and A.~Raghunathan, ``Communication-efficient
  view-pooling for distributed multi-view neural networks,'' in \emph{Proc.
  Design Automat. Test Eur. Conf. Exhib. (DATE)}, Grenoble, France, Sep. 2020.

\bibitem{choi2019context}
J.~Choi, Z.~Hakimi, P.~W. Shin, J.~Sampson, and V.~Narayanan, ``Context-aware
  convolutional neural network over distributed system in collaborative
  computing,'' in \emph{Proc. ACM/IEEE Design Automat. Conf. (DAC)}, Las Vegas,
  NV, USA, Jun. 2019.

\bibitem{Multi-Robot_Collaborative_Percep_GNN}
Y.~Zhou, J.~Xiao, Y.~Zhou, and G.~Loianno, ``Multi-robot collaborative
  perception with graph neural networks,'' \emph{IEEE Robot. Automat. Lett.},
  vol.~7, no.~2, pp. 2289--2296, Jan. 2022.

\bibitem{su2015multi_MVCNN}
H.~Su, S.~Maji, E.~Kalogerakis, and E.~Learned-Miller, ``Multi-view
  convolutional neural networks for 3d shape recognition,'' in \emph{Proc. Int.
  Conf. Comput. Vision}, Santiago, Chile, Dec. 2015.

\bibitem{alemi2016deepVIB}
A.~A. Alemi, I.~Fischer, J.~V. Dillon, and K.~Murphy, ``Deep variational
  information bottleneck,'' in \emph{Proc. Int. Conf. Learn. Repr. (ICLR)},
  Toulon, France, Apr. 2017.

\bibitem{e22090999_ceb}
I.~Fischer, ``The conditional entropy bottleneck,'' \emph{Entropy}, vol.~22,
  no.~9, Sep. 2020.

\bibitem{courtade2013multiterminal}
T.~A. Courtade and T.~Weissman, ``Multiterminal source coding under logarithmic
  loss,'' \emph{IEEE Trans. Inf. Theory}, vol.~60, no.~1, pp. 740--761, Nov.
  2013.

\bibitem{CEO_problem}
Y.~Uğur, I.~E. Aguerri, and A.~Zaidi, ``Vector gaussian ceo problem under
  logarithmic loss and applications,'' \emph{IEEE Trans, Inf. Theory}, vol.~66,
  no.~7, pp. 4183--4202, Feb. 2020.

\bibitem{DPI1_6875389}
V.~Anantharam, A.~Gohari, S.~Kamath, and C.~Nair, ``On hypercontractivity and a
  data processing inequality,'' in \emph{IEEE Int. Symp. Inf. Theory}, Jun.
  2014, pp. 3022--3026.

\bibitem{blei2017variational_survey}
D.~M. Blei, A.~Kucukelbir, and J.~D. McAuliffe, ``Variational inference: A
  review for statisticians,'' \emph{J. Amer. Statist. Assoc.}, vol. 112, no.
  518, pp. 859--877, Jul. 2017.

\bibitem{kingma2013autovae}
D.~P. Kingma and M.~Welling, ``Auto-encoding variational bayes,'' in
  \emph{Proc. Int. Conf. Learn. Repr. (ICLR)}, Banff, Canada, Apr. 2014.

\bibitem{goldfeld2020information_B_application}
Z.~Goldfeld and Y.~Polyanskiy, ``The information bottleneck problem and its
  applications in machine learning,'' \emph{IEEE J. Sel. Areas Inf. Theory},
  vol.~1, no.~1, pp. 19--38, Apr. 2020.

\bibitem{bengio2013estimating_STE}
Y.~Bengio, N.~L{\'e}onard, and A.~Courville, ``Estimating or propagating
  gradients through stochastic neurons for conditional computation,''
  \emph{arXiv preprint arXiv:1308.3432}, 2013. [Online]. Available:
  https://arxiv.org/abs/1308.3432.

\bibitem{kurka2019successive_refinement}
D.~B. Kurka and D.~G{\"u}nd{\"u}z, ``Successive refinement of images with deep
  joint source-channel coding,'' in \emph{Proc. IEEE Int. Wkshop Signal
  Process. Adv. Wireless Commun. (SPAWC)}, Cannes, France, Jul. 2019.

\bibitem{chen2020learning_to_stop}
X.~Chen, H.~Dai, Y.~Li, X.~Gao, and L.~Song, ``Learning to stop while learning
  to predict,'' in \emph{Int. Conf. Mach. Learn.}, Vienna, Australia, Apr.
  2020.

\bibitem{enomoto2021learning_to_cascade}
S.~Enomoto and T.~Eda, ``Learning to cascade: Confidence calibration for
  improving the accuracy and computational cost of cascade inference systems,''
  in \emph{Proc. AAAI Conf. Artif. Intell.}, Feb. 2021. [Online]. Available:
  https://www.aaai.org/AAAI21Papers/AAAI-2189.EnomotoS.pdf.

\bibitem{network_information_theory}
A.~El~Gamal and Y.-H. Kim, \emph{Network information theory}.\hskip 1em plus
  0.5em minus 0.4em\relax Cambridge University Press, 2011.

\bibitem{lecun1998gradientMNIST}
Y.~{Lecun}, L.~{Bottou}, Y.~{Bengio}, and P.~{Haffner}, ``Gradient-based
  learning applied to document recognition,'' \emph{Proc. IEEE}, vol.~86,
  no.~11, pp. 2278--2324, May 1998.

\bibitem{cifar10/100}
A.~Krizhevsky, G.~Hinton \emph{et~al.}, ``Learning multiple layers of features
  from tiny images,'' 2009. [Online]. Available:
  https://www.cs.toronto.edu/$\sim$kriz/learning-features-2009-TR.pdf.

\bibitem{tiny_imagenet}
P.~Chrabaszcz, I.~Loshchilov, and F.~Hutter, ``A downsampled variant of
  imagenet as an alternative to the cifar datasets,'' \emph{arXiv preprint
  arXiv:1707.08819}, 2017. [Online]. Available:
  https://arxiv.org/abs/1707.08819.

\bibitem{Neural_DSC}
J.~Whang, A.~Acharya, H.~Kim, and A.~G. Dimakis, ``Neural distributed source
  coding,'' \emph{arXiv preprint arXiv:2106.02797}, 2021. [Online]. Available:
  https://arxiv.org/abs/2106.02797.

\bibitem{WRGBD}
K.~Lai, L.~Bo, X.~Ren, and D.~Fox, ``A large-scale hierarchical multi-view
  rgb-d object dataset,'' in \emph{Proc. IEEE Int. Conf. Robot. Automat.},
  Shanghai, China, May 2011, pp. 1817--1824.

\bibitem{wu20153d_shapenet40}
Z.~Wu, S.~Song, A.~Khosla, F.~Yu, L.~Zhang, X.~Tang, and J.~Xiao, ``3{D}
  {S}hapenets: A deep representation for volumetric shapes,'' in \emph{Proc.
  IEEE/CVF Conf Comput. Vision Pattern Recognit. (CVPR)}, Boston, MA, USA, Jun.
  2015.

\bibitem{VGG}
K.~Simonyan and A.~Zisserman, ``Very deep convolutional networks for
  large-scale image recognition,'' in \emph{Int. Conf. Learn. Repr.}, 2015.

\bibitem{PreActResNet}
K.~He, X.~Zhang, S.~Ren, and J.~Sun, ``Identity mappings in deep residual
  networks,'' in \emph{Proc. Eur. Conf. Comput. Vision}, 2016, pp. 630--645.

\bibitem{zayas20173gpp_iot}
A.~D. Zayas and P.~Merino, ``The {3GPP} {NB}-{I}o{T} system architecture for
  the internet of things,'' in \emph{Proc. IEEE Int. Conf. Commun. Workshop},
  May 2017, pp. 277--282.

\bibitem{reviewer_Moldoveanu2}
M.~Moldoveanu and A.~Zaidi, ``In-network learning for distributed training and
  inference in networks,'' in \emph{Proc. IEEE GLOBECOM}, Dec. 2021.

\end{thebibliography}


\begin{thebibliography}{1}

\bibitem{IEEEhowto:kopka}
H.~Kopka and P.~W. Daly, \emph{A Guide to \LaTeX}, 3rd~ed.\hskip 1em plus
  0.5em minus 0.4em\relax Harlow, England: Addison-Wesley, 1999.

\end{thebibliography}

\iffalse

\fi

\end{document}